\documentclass[12pt]{article}

\usepackage{epsfig,amsmath,amsfonts,epsfig,amssymb,verbatim,mathrsfs}
\usepackage{hyperref,dcolumn,bm}

\newcommand{\be}{\begin{equation}}
\newcommand{\ee}{\end{equation}}
\newcommand{\ben}{\begin{equation*}}
\newcommand{\een}{\end{equation*}}
\newcommand{\bea}{\begin{eqnarray*}}
\newcommand{\eea}{\end{eqnarray*}}

\newcommand{\tev}{{\rm TeV}}
\newcommand{\gev}{{\rm GeV}}

\newcommand{\til}{\tilde}

\newcommand{\mgrav}{m_{3/2}}
\newcommand{\Ux}{U(1)_{x}}
\newcommand{\Uhyp}{U(1)_{Y}}

\newcommand{\gx}{g_{x}}
\newcommand{\Mx}{M_{x}}

\newcommand{\mhp}{m_{H}^{2}}
\newcommand{\mhm}{m_{H'}^{2}}
\newcommand{\eps}{\epsilon}

\newcommand{\stst}{\til t_{1} \bar{\til t}_{1}}
\newcommand{\sbsb}{\til b_{1} \bar{\til b}_{1}}

\newcommand{\abs}[1]{\left\lvert #1 \right\rvert}

\def\beq{\begin{eqnarray}}
\def\eeq{\end{eqnarray}}
\def\bea{\begin{eqnarray}}
\def\eea{\end{eqnarray}}

\def\tev{\, {\rm TeV}}
\def\gev{\, {\rm GeV}}

\newcommand{\gsim}{\lower.7ex\hbox{$\;\stackrel{\textstyle>}{\sim}\;$}}
\newcommand{\lsim}{\lower.7ex\hbox{$\;\stackrel{\textstyle<}{\sim}\;$}}

\def\stilde{\widetilde}

\newcommand{\newc}{\newcommand}
\newc{\Nc}{N_{c}}
\newc{\CG}{C_G}
\newc{\gp}{g'}
\newc{\stopi}{\stilde t_i}
\newc{\sboti}{\stilde b_i}
\newc{\staui}{\stilde \tau_i}
\newc{\stopj}{\stilde t_j}
\newc{\sbotj}{\stilde b_j}
\newc{\stauj}{\stilde \tau_j}
\newc{\stopI}{\stilde t_1}
\newc{\stopII}{\stilde t_2}
\newc{\sbotI}{\stilde b_1}
\newc{\sbotII}{\stilde b_2}
\newc{\stauI}{\stilde \tau_1}
\newc{\stauII}{\stilde \tau_2}
\newc{\sstop}{s_{t}}
\newc{\cstop}{c_{t}}
\newc{\ssbot}{s_{b}}
\newc{\csbot}{c_{b}}
\newc{\sstau}{s_{\tau}}
\newc{\cstau}{c_{\tau}}
\newc{\Sstop}{s_{2t}}
\newc{\Cstop}{c_{2t}}
\newc{\Ssbot}{s_{2b}}
\newc{\Csbot}{c_{2b}}
\newc{\Sstau}{s_{2\tau}}
\newc{\Cstau}{c_{2\tau}}
\newc{\salpha}{s_\alpha}
\newc{\calpha}{c_\alpha}
\newc{\Calpha}{c_{2\alpha}}
\newc{\Salpha}{s_{2\alpha}}
\newc{\sbetapm}{s_{\beta_\pm}}
\newc{\cbetapm}{c_{\beta_\pm}}
\newc{\Sbetapm}{s_{2 \beta_\pm}}
\newc{\Cbetapm}{c_{2 \beta_\pm}}
\newc{\sbetaO}{s_{\beta_0}}
\newc{\cbetaO}{c_{\beta_0}}
\newc{\SbetaO}{s_{2 \beta_0}}
\newc{\CbetaO}{c_{2 \beta_0}}
\newc{\vu}{v_u}
\newc{\vd}{v_d}
\newc{\seL}{\stilde e_L}
\newc{\smuL}{\stilde \mu_L}
\newc{\seR}{\stilde e_R}
\newc{\smuR}{\stilde \mu_R}
\newc{\suL}{\stilde u_L}
\newc{\sdL}{\stilde d_L}
\newc{\suR}{\stilde u_R}
\newc{\sdR}{\stilde d_R}
\newc{\scL}{\stilde c_L}
\newc{\ssL}{\stilde s_L}
\newc{\scR}{\stilde c_R}
\newc{\ssR}{\stilde s_R}
\newc{\snue}{\stilde \nu_e}
\newc{\snumu}{\stilde \nu_\mu}
\newc{\snutau}{\stilde \nu_\tau}
\newc{\Gpm}{G^\pm}
\newc{\Hpm}{H^\pm}
\newc{\FFbS}{\overline{FF}S}
\newc{\FFbV}{\overline{FF}V}
\newc{\FSS}{F_{SS}}
\newc{\FSSS}{F_{SSS}}
\newc{\FFFS}{F_{FFS}}
\newc{\FFFbS}{F_{\overline{FF}S}}
\newc{\FSSV}{F_{SSV}}
\newc{\FVS}{F_{VS}}
\newc{\FVVS}{F_{VVS}}
\newc{\FFFV}{F_{FFV}}
\newc{\FFFbV}{F_{\overline{FF}V}}
\newc{\Fgauge}{F_{\rm gauge}}
\newc{\DRbarprime}{$\overline{\rm DR}'$ }
\newc{\DRbar}{$\overline{\rm DR}$ }
\newc{\MSbar}{$\overline{\rm MS}$ }
\newc{\Yu}{{\bf Y}_u}
\newc{\Yd}{{\bf Y}_d}
\newc{\Ye}{{\bf Y}_e}
\newc{\Au}{{\bf a}_u}
\newc{\Ad}{{\bf a}_d}
\newc{\Ae}{{\bf a}_e}
\newc{\zhol}{Z^{\rm hol}}
\newc{\ra}{\rightarrow}
\newc{\ccdot}{\!\cdot\!}

\newcommand{\nnmb}{\nonumber}

\newcommand{\lrf}[2]{\left(\frac{#1}{#2}\right)}

\begin{document}

\setlength{\baselineskip}{0.2in}


\begin{titlepage}
\noindent
\vspace{1cm}

\begin{center}
  \begin{Large}
    \begin{bf}
Kinetically-Enhanced Anomaly Mediation
     \end{bf}
  \end{Large}
\end{center}
\vspace{0.2cm}

\begin{center}

\begin{large}
Abhishek Kumar, David E. Morrissey, and Andrew Spray\\
\end{large}
\vspace{1cm}
  \begin{it}
TRIUMF,
4004 Wesbrook Mall, Vancouver, BC V6T 2A3, Canada.
\vspace{0.5cm}\\
email: abhishek@triumf.ca, dmorri@triumf.ca, aps37@triumf.ca
\vspace{0.5cm}
\end{it}

\end{center}

\center{\today}

\begin{abstract}

  We investigate a modification of anomaly-mediated supersymmetry
breaking~(AMSB) with an exotic $U(1)_x$ gauge sector that
can solve the tachyonic slepton problem of minimal AMSB scenarios.
The new $U(1)_x$ multiplet is assumed to couple directly to the
source of supersymmetry breaking, but only indirectly to the 
 minimal supersymmetric Standard Model~(MSSM) through
kinetic mixing with hypercharge.  If the MSSM sector is also sequestered
from the source of supersymmetry breaking, the contributions to
the MSSM soft terms come from both AMSB and the $U(1)_x$ kinetic 
coupling.  We find that this arrangement can give rise to a 
flavour-universal, phenomenologically viable, and distinctive spectrum 
of MSSM superpartners.
We also investigate the prospects for discovery and the most likely
signatures of this scenario at the Large Hadron Collider~(LHC).

\end{abstract}

\vspace{1cm}

\end{titlepage}

 \setcounter{page}{2}





\section{Introduction}

 Supersymmetry is one of the leading candidates for new physics
beyond the Standard Model~(SM).  To be consistent with existing experimental
searches, the SM superpartners must have supersymmetry-breaking masses
near the electroweak scale.  The underlying nature of supersymmetry
breaking plays a central role in determining the mass spectrum 
and the interactions of these new superpartner states, 
and therefore also the predictions of the theory at particle colliders.  
The current LHC era holds the promise of discovering superparticles 
and might also shed light on the nature of supersymmetry breaking. 
Thus it is all the more important to focus on supersymmetry breaking 
and to understand its origin.

  A variety of mechanisms have been proposed to explain the breaking
of supersymmetry and its mediation to the SM superpartners.
The leading candidates include gravity mediation,
gauge mediation, and anomaly mediation.  Anomaly-mediated supersymmetry
breaking~(AMSB) is particularly compelling because it avoids introducing
too much flavour mixing to the superpartner interactions and it relies
only on fields present in the minimal supergravity
multiplet~\cite{Randall:1998uk,Giudice:1998xp}.
Unfortunately, AMSB in its simplest form also predicts negative soft
squared masses for the sleptons that would induce an unacceptable
spontaneous breakdown of electromagnetism.  To remedy the situation,
many solutions have been proposed to generate a viable slepton spectrum
such as non-decoupling thresholds that produce additional 
gauge mediation~\cite{Pomarol:1999ie,Katz:1999uw,Sundrum:2004un,Cai:2010tj,Kobayashi:2011bp,Okada:2002mv},
new $D$-term contributions to the scalar 
masses~\cite{Jack:2000cd,ArkaniHamed:2000xj},
unsequestered vector multiplets~\cite{Kaplan:2000jz,Chacko:2001jt,
Dermisek:2007qi,deBlas:2009vx},
new couplings involving the lepton multiplets that persist to low
energies~\cite{Chacko:1999am,Mohapatra:2007xq}, 
and moderate additional gravity-mediated contributions~\cite{Choi:2005ge,
Endo:2005uy,Acharya:2007rc}.

\begin{figure}[ttt]
\begin{center}
\includegraphics[width=0.4\textwidth,viewport={100 550 450 770}]{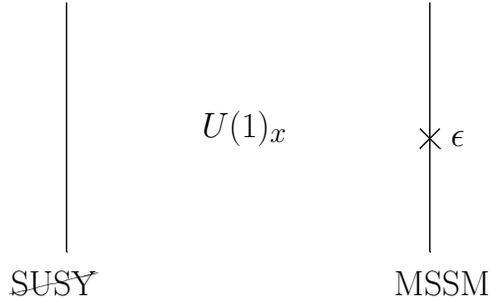}
\caption{Schematic diagram of the visible, $U(1)_x$,
and supersymmetry-breaking sectors.
\label{fig:AMSBbranes}}
\end{center}
\end{figure}

  In the present work we investigate a new mechanism for fixing the
negative slepton squared mass problem of AMSB that relies on supersymmetric
kinetic mixing between an unsequestered $U(1)_x$ gauge symmetry and hypercharge.
It is similar to but distinct from the approaches used 
in Refs.~\cite{Dermisek:2007qi,deBlas:2009vx}.
A schematic diagram of our proposal is shown in Fig.~\ref{fig:AMSBbranes}.
As in standard AMSB, the minimal supersymmetric extension of the Standard
Model~(MSSM) is \emph{sequestered} from the source of
supersymmetry breaking.  This can potentially arise from a localization of the
corresponding fields in different regions in an extra
dimension~\cite{Randall:1998uk,Luty:1999cz,Kachru:2007xp}
or through the effects of conformal running~\cite{Luty:2001zv,Schmaltz:2006qs}.
Despite the sequestering, supersymmetry breaking is still communicated 
to the MSSM through universal interactions with the supergravity multiplet.  
We augment this minimal picture with an exotic $U(1)_x$ gauge sector that
is not sequestered from supersymmetry breaking and that therefore
receives large direct contributions to its soft spectrum from standard
gravity mediation.  The additional supersymmetry breaking present
in the $U(1)_x$ sector is then communicated to the MSSM through
gauge kinetic mixing with hypercharge, modifying the mass spectrum
of MSSM superpartners and potentially solving the problem of
negative slepton squared masses.

  The outline of this paper is as follows.  In Section~\ref{sec:setup}
we describe in more detail the underlying extension of the MSSM
and its effect on the mass spectrum of the SM superpartners.
Next, in Section~\ref{sec:viable} we show that these modifications can
lead to a phenomenologically viable spectrum.  In Section~\ref{sec:coll} we
investigate the collider signatures of this scenario, while
Section~\ref{sec:conc} is reserved for our conclusions.  Some accompanying
technical details can be found in Appendices~\ref{sec:apprg} and \ref{sec:ft}.

\section{Couplings and Soft Terms\label{sec:setup}}

  Our theory consists of the MSSM together with an exotic supersymmetric
$x$ sector.  The $x$ sector contains a $U(1)_x$ gauge multiplet,
and a pair of chiral multiplets $H$ and $H'$ with equal and opposite 
$U(1)_x$ charges $\pm x_H$ and superpotential
\beq
W_{x} = \mu'HH' \ .
\label{eq:xsup}
\eeq
The $x$ and visible sectors couple only via supersymmetric gauge kinetic mixing,
\beq
\mathscr{L} \supset \int d^2\theta
\left(\frac{\eps}{2} B^{\alpha}X_{\alpha}
+ \frac{1}{4} X^{\alpha}X_{\alpha}
+ \frac{1}{4} B^{\alpha}B_{\alpha}\right) + \rm h.c.
\label{eq:kinmix}
\eeq
where $B_{\alpha}$ and $X_{\alpha}$ are the gauge field strengths of
$\Uhyp$ and $\Ux$ respectively.  This mixing interaction can be
generated by integrating out massive chiral multiplets charged under both
gauge groups~\cite{Holdom:1985ag} giving rise to natural values
of $\epsilon \sim 10^{-4}\!-\!10^{-2}$.  We treat the kinetic mixing
coupling as a free parameter with the only condition that $\epsilon \ll 1$.

\subsection{Supersymmetry Breaking and Sequestering}

  We assume that the MSSM sector is sequestered from the source of
supersymmetry breaking, but that the $U(1)_x$ sector is not.
This can be realized by localizing the MSSM away from
supersymmetry breaking in a warped extra-dimensional
setup~\cite{Randall:1998uk,Luty:1999cz,Kachru:2007xp},
or through the effects of conformal
running~\cite{Luty:2001zv,Schmaltz:2006qs}.
We assume further that the source of gauge kinetic
mixing is also sequestered.  A schematic picture of the sequestering is given
in Fig.~\ref{fig:AMSBbranes}.

  In this framework the dominant source of supersymmetry breaking
in the $x$ sector comes from direct gravity mediation.
Typical soft terms in the $x$ sector are therefore
\beq
M_x \sim \sqrt{|m_H^2|} \sim \sqrt{|m_{H'}^2|} \sim \sqrt{b'} \sim m_{3/2},
\eeq
where $M_x$ is the $U(1)_x$ gaugino soft mass, $m_{H,H'}^2$ are the soft
squared masses of $H$ and $H'$, $b'$ is the bilinear holomorphic
soft term mixing $H$ and $H'$, and $m_{3/2}$ is the gravitino mass.
These soft masses are not calculable without an ultraviolet completion
incorporating supergravity and we treat them as free parameters
on the order of $m_{3/2}$.\footnote{
For $M_x,\,\sqrt{b'} \sim m_{3/2}$ it is also necessary that the
source of supersymmetry breaking involves gauge singlets.}
The dimensionful supersymmetric coupling $\mu'$ appearing
in Eq.~\eqref{eq:xsup} can also be generated with $\mu'\sim m_{3/2}$
via the Giudice-Masiero mechanism~\cite{Giudice:1988yz},
as can the corresponding soft coupling $b'$.
Negative values of $m_{H}^2$ or $m_{H'}^2$ can induce the spontaneous
breakdown of $U(1)_x$ with natural values of $\langle H \rangle \sim
\langle H'\rangle \sim m_{3/2}/g_x$, leading to a massive $Z_x$ vector
of mass $m_{x} \sim m_{3/2}$.

  Soft supersymmetry breaking terms in the MSSM sector
will be generated by anomaly mediation together with
direct mediation from the $x$ sector via gauge kinetic mixing.
As such, they will be suppressed relative to the soft breaking parameters
in the $x$ sector due to sequestering and the small value of
the kinetic mixing coupling $\epsilon$.  We present a prescription
to compute the MSSM soft terms below, and we give a derivation of
this prescription in Appendix~\ref{sec:apprg}.

\subsection{MSSM Soft Masses}

  Supersymmetry breaking couplings in the MSSM receive contributions 
from anomaly mediation and from gauge kinetic mixing with the unsequestered 
$x$ sector.  In computing these soft terms, we take $\Lambda$ to be the 
characteristic mass scale of the sequestering dynamics, such as the 
compactification scale of an extra dimension or the exit scale from 
a period of conformal running.  With an eye on grand unification, 
we choose $\Lambda = 2\times 10^{16}\,\gev$,
but other values could be considered.  For simplicity, we also identify
$\Lambda$ with the (sequestered) dynamics giving rise to gauge kinetic
mixing, such as a vector-like pair of chiral multiplets charged under
both $U(1)_x$ and hypercharge with a supersymmetric mass on the order
of $\Lambda$.

\begin{figure}[ttt]
\begin{center}
\includegraphics[width=0.4\textwidth,viewport={100 550 500 770}]{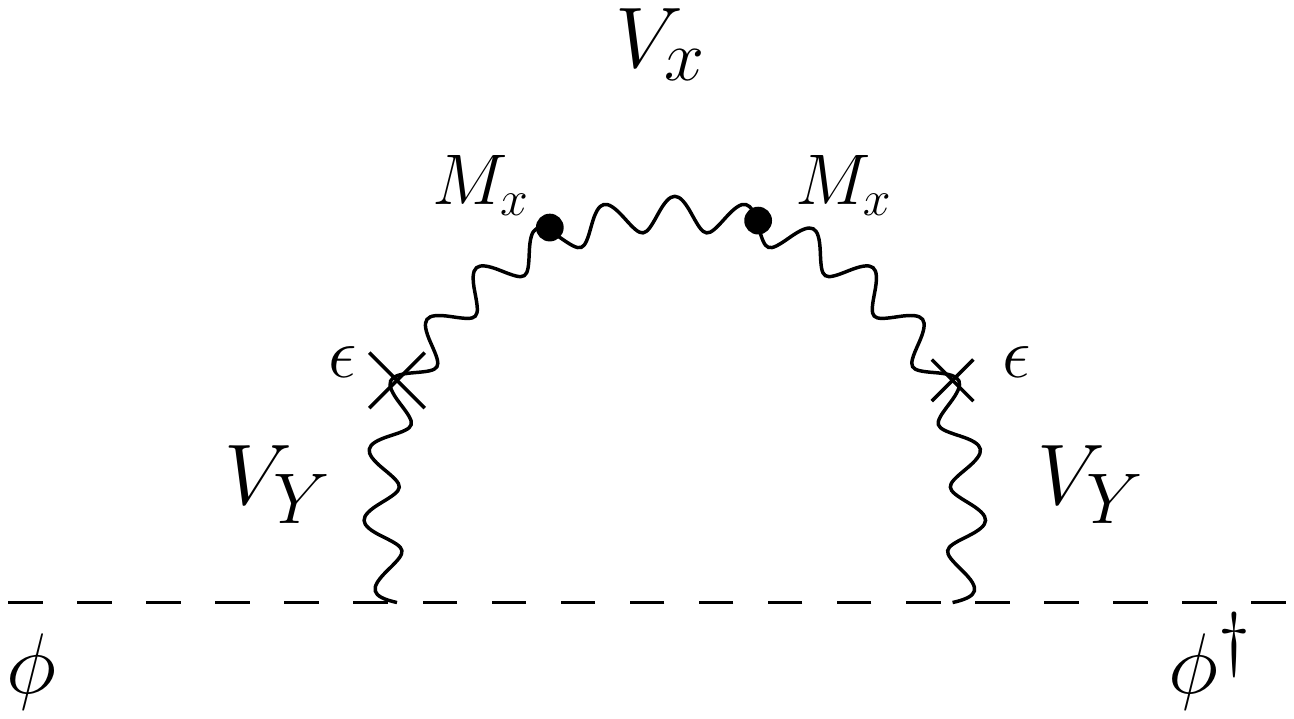}
\includegraphics[width=0.4\textwidth,viewport={100 425 500 760}]{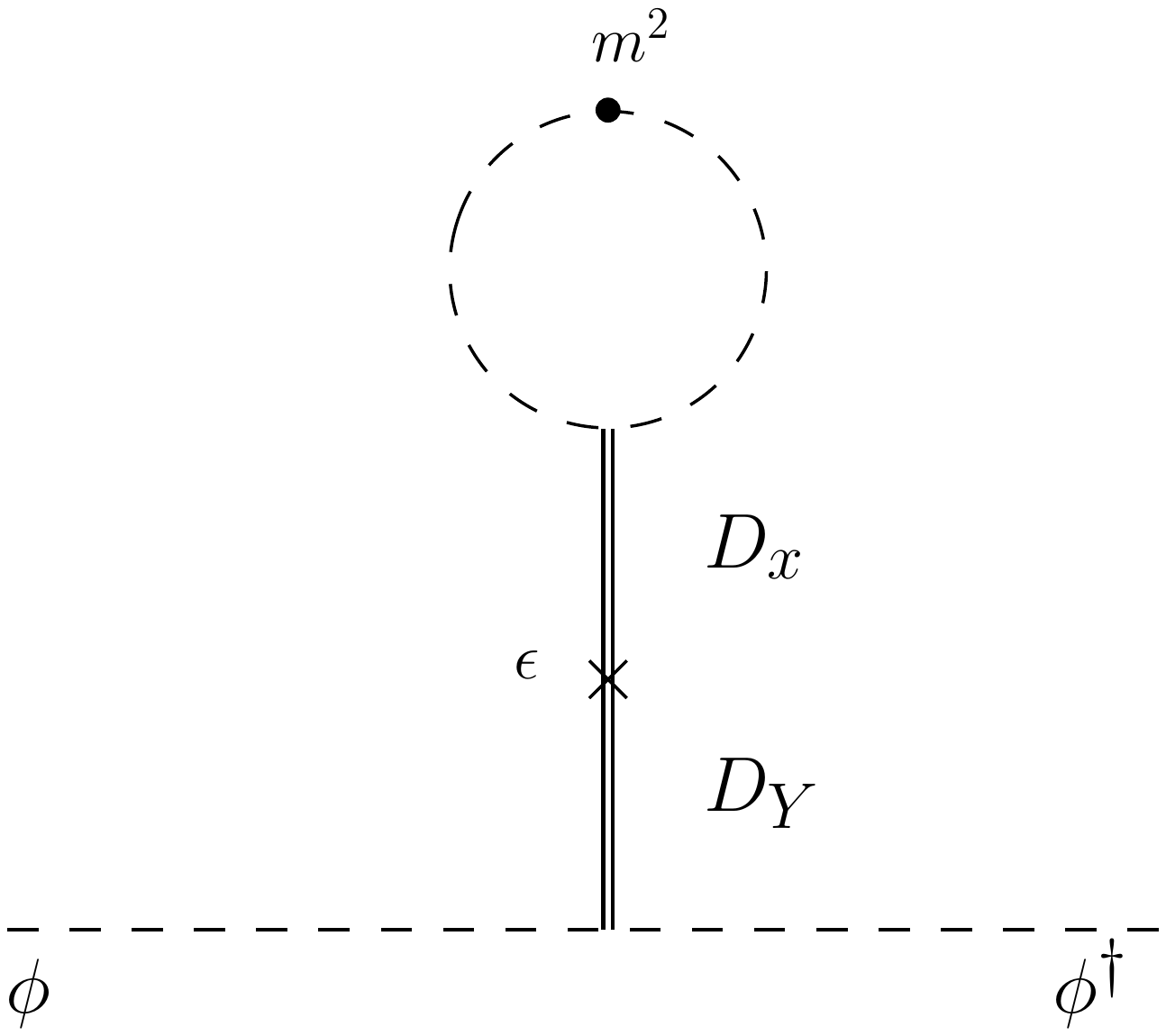}
\caption{Contributions to the RG running of the MSSM scalar soft masses from
supersymmetric gauge kinetic mixing between hypercharge and $U(1)_x$.
The left panel shows the leading effect from the mixing of the gaugino
components, while the right panel illustrates the contribution from the
mixing of the $D$-terms.}
\label{fig:Mxloop}
\label{fig:Sterm}
\end{center}
\end{figure}

  At scale $\Lambda$, the MSSM soft masses are dominated by the usual
contributions from anomaly mediation evaluated at that scale,
given explicitly in Refs.~\cite{Randall:1998uk,Luty:1999cz,Gherghetta:1999sw}.
We take this to be our high-energy boundary condition.
Additional threshold corrections as well as the anomaly-mediated
contribution of the gauge kinetic mixing interaction are parametrically
smaller.
In the course of renormalization group~(RG) evolution down from $\Lambda$,
the MSSM soft masses receive additional contributions from the $x$ sector.
At one-loop order the MSSM gaugino masses remain unchanged.  For the
scalar soft masses, we find (to quadratic order in $\epsilon \ll 1$)
\bea
(4\pi)^2\frac{dm_i^2}{dt} &=& (4\pi)^2\lrf{dm_i^2}{dt}_{\!MSSM}
-\frac{24}{5}Y_i^2\epsilon^2g_1^2|M_x|^2
-2\sqrt{\frac{3}{5}}Y_ig_1g_x\epsilon\,S_x
\label{misoft}
\eea
where $t = \ln(Q/m_Z)$ is the logarithm of the renormalization scale $Q$,
$S_x = x_H(m_H^2-m_{H'}^2)$, and $M_x$ is the $x$-sector gaugino mass.
The first new contribution is analogous to gaugino 
mediation~\cite{Kaplan:1999ac,Chacko:1999mi} 
and is generated by diagrams of the form shown in the left panel 
of Fig.~\ref{fig:Mxloop},
while the second comes from diagrams like in the right panel of this figure.
The one-loop running of the trilinear $A$ terms is given by
(to quadratic order in $\epsilon \ll 1$)
\beq
(4\pi)^{2} \frac{dA_{f}}{dt} = (4\pi)^2 \lrf{dA_{f}}{dt}_{MSSM}
+ \frac{12}{5}\sum_iY_i^2\,g_{1}^{2}\eps^{2}M_{x},
\label{aisoft}
\eeq
where the sum in the second term runs over the hypercharges of all
the fields involved in the corresponding interaction.
This contribution is generated by diagrams of the form of
Fig.~\ref{fig:Mxloop}, but with only a single $M_x$ insertion.
RG equations for $M_x$ and $m_{H,H'}^2$ are given in Appendix~\ref{sec:apprg}.

  These RG equations only apply at energies larger than the $x$-sector
masses, on the order of $m_{3/2}$.  At $Q = m_{3/2}$ we integrate out
the heavy $x$-sector multiplets.  The main effect of doing so is to generate
an effective Fayet-Iliopoulos~(FI) term~\cite{Fayet:1974jb} for hypercharge 
from the vacuum expectation values (VEVs) of $H$ and $H'$ inducing 
the spontaneous breakdown of $U(1)_x$.  Writing
\beq
\langle H\rangle = v_x\,\sin\alpha,~~~\langle H'\rangle = v_x\,\cos\alpha \ ,
\eeq
we have $m_x = \sqrt{2}g_xc_{\epsilon}x_Hv_x$ for the $U(1)_x$ vector mass,
where $c_{\epsilon} = 1/\sqrt{1-\epsilon^2}$.
Up to an overall constant, the induced FI term can be absorbed in
the MSSM soft scalar masses according to (to quadratic order in $\epsilon\ll 1$)
\beq
m_i^2(Q=m_{3/2}^-) = m_i^2(Q=m_{3/2}^+)
+ \frac{\epsilon}{2}\sqrt{\frac{3}{5}}\frac{Y_i}{x_H}\frac{g_1}{g_x}
m_x^2\cos2\alpha \ .
\label{dshift}
\eeq
The RG running of the MSSM soft terms from $m_{3/2}$ to
near the electroweak scale is identical to the MSSM.
More details about the prescription given here for computing the
MSSM soft masses are given in Appendix~\ref{sec:apprg}.

  From the discussion above we see that, in addition to anomaly mediation,
the MSSM soft scalar masses get contributions from the $U(1)_x$ gauginos
that push them towards more positive values, as well as contributions
from $S_x$ and $U(1)_x$ breaking that are proportional to hypercharge
and can have either sign.  We will show below that the positive
effect of the unsequestered $U(1)_x$ gaugino mass can yield a
phenomenologically viable soft mass spectrum for the MSSM.
In particular, this scenario can evade the tachyonic slepton problem
of minimal anomaly mediation.

  There are two additional important differences between this scenario and
pure anomaly mediation.  First, the MSSM soft masses depend implicitly
on high-scale physics, and are therefore not UV insensitive.
This is the result of the overall mass scale in the $x$ sector being
of the same order as the scale of supersymmetry breaking in that sector.
Second, to get $M_x \sim m_{3/2}$, a necessary condition for positive MSSM
soft masses, the supersymmetry breaking sector must contain gauge singlets,
which is not the case for pure anomaly mediation.
 
  A slight variation on this scenario that could lead to positive
soft masses for all MSSM scalars in the absence of a supersymmetry-breaking
singlet (or for small values of $\epsilon$) would be to gauge a sequestered 
$(B\!-\!L)$ and to arrange for it to also have kinetic mixing with $U(1)_X$.  
In this case, both hypercharge and $(B\!-\!L)$ $D$-terms would be generated  
by the $x$ sector.  As shown in Refs.~\cite{Jack:2000cd,ArkaniHamed:2000xj}, 
such terms can yield positive MSSM soft scalar masses while retaining the 
UV insensitivity property of minimal AMSB.  We defer a study of this
variation to a future work.

\section{$X$-Sector Parameters and MSSM Spectra
\label{sec:viable}}

    Based on our discussion in the previous section, we can derive the
low-energy MSSM spectrum once we specify the parameter values in the
$x$ sector (along with $\mu$ and $B\mu$ in the MSSM sector).
While the $x$ sector is relatively unconstrained, there are evidently
many phenomenological constraints on the visible sector, such as
consistent electroweak symmetry breaking, precision electroweak tests,
and direct collider constraints on particle masses.  In this section
we perform a scan over parameter values in the $x$ sector to search
for viable MSSM spectra.

\subsection{Symmetry Breaking in the $X$ Sector}

  The independent parameters characterizing the $x$ sector are
$\epsilon$, $g_xx_H$, $\mu'$, $b'$, $M_x$, $m_{H}^2$,
and $m_{H'}^2$.
Of these, the first three are supersymmetric while the remaining
four arise from soft supersymmetry breaking.
We demand that the $U(1)_x$ gauge symmetry be spontaneously
broken.  The corresponding minimization conditions for $H$
and $H'$ allow us to replace two of these parameters with
the massive $Z_x$ vector mass $m_x$ and the ratio of VEVs $\tan\alpha$.
We choose to solve for $b'$, $m_H^2$, and $m_{H'}^2$ in terms
of $m_x$, $\tan\alpha$, and $S_x/x_H = (m_H^2-m_{H'}^2)$.

  The leading part of the $x$-sector scalar potential is
\beq
V = (|\mu'|^{2} + \mhp)|H|^{2} + (|\mu'|^{2} + \mhm)|H'|^{2}
+ \frac{1}{2} \gx^{2}c_{\epsilon}^2\left(|H|^{2}-|H'|^{2}\right)^{2}
- \left(b' H H'+h.c.\right)
\label{eq:xpot}
\eeq
We can take $b'$ to be real and positive by making a field redefinition,
in which case the potential is minimized with $H$ and $H'$ both
real and positive as well.
Minimizing the potential gives the relations
\bea
b' &=& -\frac{1}{2} m_{x}^{2} \sin(2\alpha)
+ \frac{1}{2} \frac{S_{x}}{x_H} \tan(2\alpha),
\label{eq:b}\\
\mhp &=& -\frac{1}{2} m_{x}^{2} - |\mu'|^{2}
+ \frac{S_{x}}{x_H} \frac{\cos^2\alpha}{\cos(2\alpha)},
\label{eq:mhp} \\
\mhm &=& -\frac{1}{2} m_{x}^{2} - |\mu'|^{2}
+ \frac{S_{x}}{x_H} \frac{\sin^2\alpha}{\cos(2\alpha)}.
\label{eq:mhm}
\eea
Corrections to these relations from the MSSM backreaction
are suppressed by powers of $\epsilon\,m_Z^2/m_x^2$ and can
be safely neglected (away from $\tan\alpha=1$).

\subsection{Scans and Experimental Constraints}

  To investigate the effect of the $x$ sector on the MSSM spectrum,
we scan over this sector by choosing the values of $M_x$, $g_xx_H$,
and $\epsilon$ at the high input scale $\Lambda$, as well as the values
of $S_x/x_H$, $m_x$, $\tan\alpha$, and $\mu'$ at the lower scale $m_{3/2}$.
Since the supersymmetry-breaking terms in the $x$ sector in this scenario
get direct gravity-mediated contributions, this is the natural scale
for these parameters as well as for the spontaneous breakdown of $U(1)_x$,
and we incorporate this property in our numerical analysis.
The scan ranges considered are listed in Table~\ref{tab:highlowscale}.

\begin{table}
  \centering
  \begin{minipage}{0.45\textwidth}
    \centering
    \begin{tabular}{|c|c|}
      \hline
      Parameter & Range \\
      \hline
      $\eps$ & [0.01,0.1] \\
      $\gx x_H$ & [0.7,0.9] \\
      $\Mx$ & [0.5,5]$\times\mgrav$ \\
      \hline
    \end{tabular}
  \end{minipage}
  \begin{minipage}{0.45\textwidth}
    \centering
    \begin{tabular}{|c|c|}
      \hline
      Parameter & Range \\
      \hline
      $S_{x}/x_H$ & [-5,5]$\times\mgrav^{2}$ \\
      $m_{x}$ & [0.1,10]$\times\mgrav$  \\
      tan $\alpha$ & [0,0.75] $\cup$[1.25,50] \\
      $\mu'$ & [0.5,5]$\times\mgrav$ \\
      \hline
    \end{tabular}
  \end{minipage}
  \caption{Input parameters in the $x$ sector specified at the high scale
$\Lambda$ (left) and the lower scale $m_{3/2}$ (right), and the
corresponding scan ranges considered in our analysis.}
  \label{tab:highlowscale}
\end{table}

  A phenomenologically viable particle spectrum must have
a spontaneously broken $U(1)_x$ gauge symmetry and an acceptable
MSSM particle spectrum.  For a given set of $x$-sector input values,
we calculate the $x$- and MSSM-sector particle masses using a numerical
code to integrate the modified one-loop RG equations based on
Eqs.~(\ref{misoft},\ref{aisoft}).  The MSSM soft masses at the 
TeV scale are then put into SuSpect~2.41~\cite{Djouadi:2002ze}, 
which calculates the physical spectrum
including threshold corrections.  In doing so, we also implicitly
fix the values of $\mu$ and $B\mu$ to achieve electroweak symmetry
breaking~(EWSB) for a given value of $\tan\beta$.
SuSpect also performs a number of checks on the spectrum, 
such as consistent EWSB and
the absence of dangerous charge or color breaking minima.
Furthermore, it calculates SUSY contributions to several
precision observables including the $\rho$ parameter,
the anomalous magnetic moment of the muon, 
and BR($b\rightarrow s\gamma$).

  We also demand that the MSSM superpartner spectrum has a neutral lightest
superpartner state~(LSP), and is consistent with existing collider bounds
from LEP and the Tevatron. Constraints and
discovery prospects from the LHC will be discussed in more
detail in Section~\ref{sec:coll}.
In particular, the next-to-LSP~(NLSP) is typically a chargino, and the LEP
bound of $m_{\chi^{\pm}} \gtrsim 95~\gev$~\cite{Heister:2002mn,Abbiendi:2003sc}
forces $m_{3/2} \gtrsim 35\,\tev$.
We also impose only a very weak bound on the lightest Higgs boson mass
of $m_h > 110\,\gev$ to allow for a theoretical uncertainty due to
higher-order corrections as well as the possibility of additional
contributions to its mass beyond those of the MSSM~\cite{
Batra:2003nj,Barger:2006dh,Martin:2009bg}.

\subsection{Viable MSSM Spectra}

  Upon computing the MSSM particle masses in this scenario, we find
superpartner spectra that are consistent with existing experimental
constraints.  Most importantly, the contributions of the $x$ sector
to the MSSM can solve the tachyonic slepton problem of minimal anomaly
mediation.  The resulting spectra are also fairly distinctive,
and could be probed in upcoming LHC searches.

  The gaugino soft masses in this scenario are identical (at one loop)
to their values in minimal anomaly mediation.  The Wino mass $M_2$ is the
smallest of these soft masses, followed by the Bino mass $M_1$ and the
gluino mass $M_3$.  Their ratio is approximately
$M_{1}:M_{2}:M_{3} \sim 3.3 : 1.0 : 7.8$.

  Scalar soft squared masses are significantly shifted by the $x$ sector.
For the first and second generations, where the Yukawa couplings can be ignored
to a good approximation, the net one-loop effect can be expressed in the form
\beq
m_i^2(Q) = [m_i^2(Q)]_{AMSB} + Y_i^2A + Y_i\,B(Q) \ ,
\label{eq:AB}
\eeq
where $Q < m_{3/2}$ is the RG running scale, $[m_i^2(Q)]_{AMSB}$ is the
value that the soft mass would have in minimal anomaly mediation,
and $A$ and $B$ are universal and given by (to quadratic order in $\epsilon$)
\bea
A &=&
\frac{1}{(4\pi)^2}\frac{24}{5}
\left(\epsilon^2g_1^2|M_x|^2\;\frac{1}{g_x^6g_1^4}\right)_{\!\!\Lambda}\:
\int_0^{\ln(\Lambda/m_{3/2})}\!\!\!dt\; g_x^6g_1^4 \ ,
\label{eq:ARG}
\\
B&=&
\frac{1}{2}\sqrt{\frac{3}{5}}\,g_1^2(Q)\,
\lrf{\epsilon}{g_1g_xx_H}
\left(m_x^2\cos({2\alpha})
+
[S_x(\Lambda)/x_H]\left[1-\frac{g_x^2(m_{3/2})}{g_x^2(\Lambda)}\right] 
\right) \ .
\label{eq:BRG}
\eea
The $A$ coefficient arises from the the heavy $U(1)_x$
gaugino mass and is clearly non-negative.  The $B$ coefficient can have either
sign, and can be related to induced Fayet-Iliopoulos terms as discussed
in Appendix~\ref{sec:apprg}.\footnote{Note that $\epsilon/g_xg_1$ 
is RG-invariant at one-loop.}
The soft scalar masses of the third-generation cannot be written in such
a simple form due to the effect of the larger Yukawa couplings on the RG
running.  Even so, the expressions in
Eqs.~(\ref{eq:AB},\ref{eq:ARG},\ref{eq:BRG}) often still provide a useful
approximation to their values.

  It is convenient to describe the effects
of the $x$ sector on the MSSM and the corresponding phenomenologically
viable regions in terms of the $A$ and $B$ coefficients.
Indeed, any set of $x$-sector parameters maps onto a unique region in
the $A$-$B$ plane.  Furthermore, it can be shown that for fixed $m_{3/2}$,
any two sets of $x$-sector parameters that map to the same point in the
$A$-$B$ plane also yield identical values for the third-generation soft
scalar masses and trilinear $A$ terms~\cite{ArkaniHamed:2000xj,Jack:2000nm}.  
Therefore any point in the $A$-$B$ plane corresponds to a unique MSSM spectrum
in this scenario for a given value of $m_{3/2}$.

\begin{figure}[ttt]
\begin{center}
\includegraphics[width=0.5\textwidth]{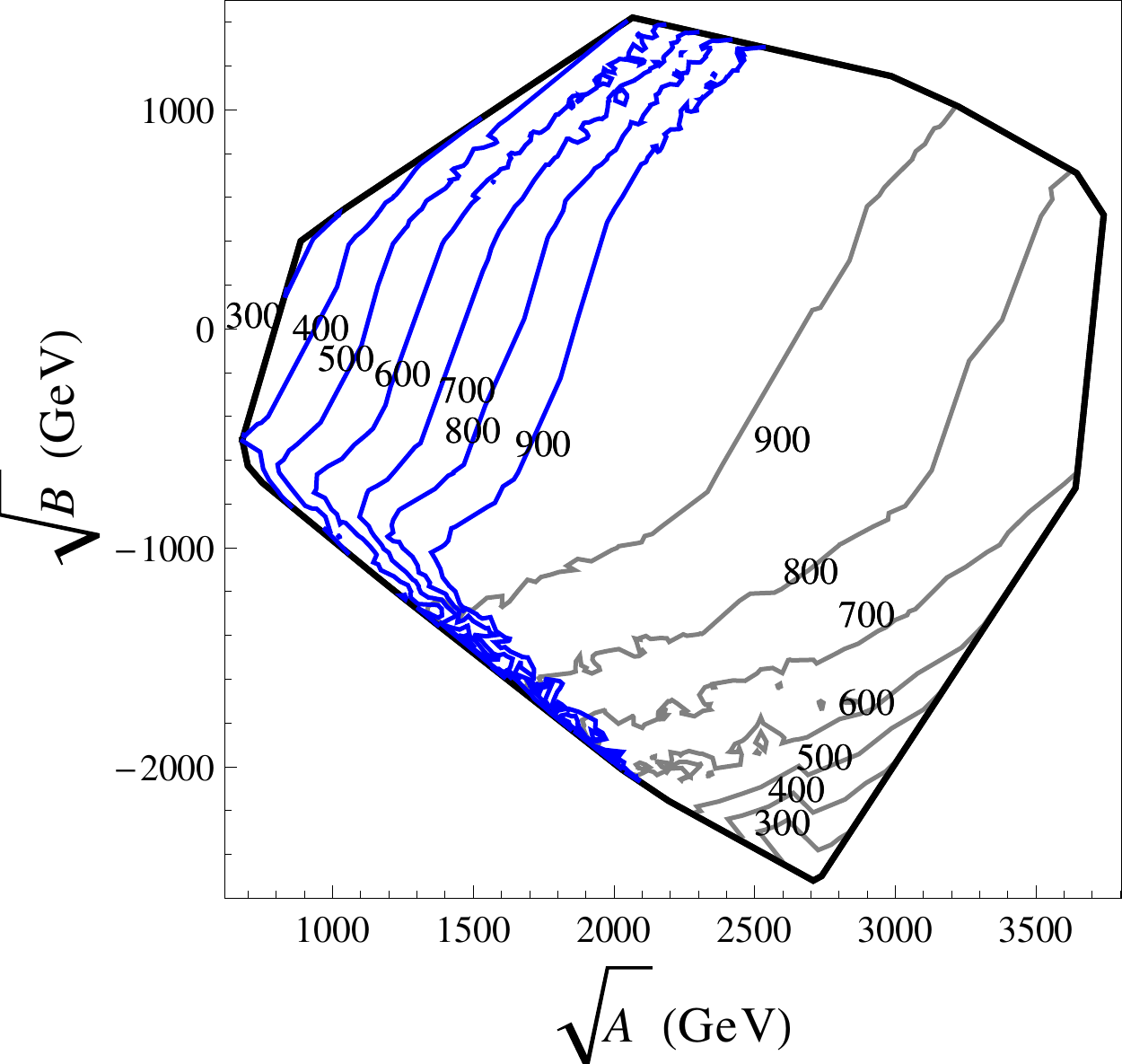}
\caption{Lightest squark (light grey) and slepton (dark blue) masses in 
the phenomenologically allowed region of the ${A}$-${B}$ plane
for $m_{3/2}=60\,\tev$. 
The lightest squarks are $\til b_{1}, \til t_{1}$. The lightest sleptons 
are $\til \nu_{\tau}$ along the south-east gradient and $\til \tau_{1}$ 
along the north-east.}
\label{fig:lightestsqkslep}
\end{center}
\end{figure}

  In Fig.~\ref{fig:lightestsqkslep} we show the allowed portion of
the $A$-$B$ plane for $m_{3/2} = 60\,\tev$.  Throughout this plane
the gaugino spectrum is identical to minimal AMSB leading to
nearly-degenerate Wino-like chargino and neutralino states with
masses close to $165\,\gev$, a mostly-Bino neutralino with mass
of about $545\,\gev$, and a $1.3\,\tev$ gluino.  Raising $m_{3/2}$
increases these masses proportionally.
On the other hand, masses of the scalar superpartners vary significantly
throughout the parameter region.  In Fig.~\ref{fig:lightestsqkslep} we
also show the mass contours of the lightest slepton ($\tilde{\nu}_\tau$ 
or $\tilde{\tau}_1$ in dark blue) and the lightest squark 
($\tilde{b}_1$ or $\tilde{t}_1$ in light grey).

  The allowed region in the $A$-$B$ plane is cut off at small values 
of $A$ by unacceptably low values of the slepton masses corresponding 
to $m^2_{L_3}$ and $m^2_{E^c_3}$.
In particular, since $Y_{L} = - \frac{1}{2}$ and $Y_{E^c} = 1$,
positive $B$ values push $m_{E^c}^2$ up and $m_{L}^2$ down.
Thus the region of small $A$ with $B>0$ is constrained by small
$m^2_{L_3}$, and small $A$ with $B<0$ is cut off by small $m^2_{E^c_3}$.
The soft masses $m^2_{U^c_3}$ and $m^2_{D^c_3}$ also scale similarly,
since  $Y_{U^c} = - \frac{2}{3}$ and $Y_{D^c} = \frac{1}{3}$.
Increasing $A$ tends to increase the values of nearly all the soft
scalar masses.  An important exception is $m^2_{Q_3}$, which runs small
due to the backreaction from the large top Yukawa contribution
in its RG equation, particularly when $B< 0$.   This leads to the
large $A$ region being constrained by phenomenological bounds.
In the region of large $A$ and $B<0$, the mass difference between
the light stops and sbottoms ($\til t_{L}$ and $\til b_{L}$) becomes
large and is constrained by the $\rho$ parameter.  For large $A$ and $B>0$,
the up-type Higgs soft mass becomes too large and positive to allow
electroweak symmetry breaking. Thus, the $\rho$ parameter and
consistent EWSB cut off the large $A$ region for $B<0$ and $B>0$
respectively.

  The allowed region in Fig.~\ref{fig:lightestsqkslep} also shows 
that we need $\sqrt{|B|} \lesssim \sqrt{A}$ to obtain a viable 
MSSM spectrum.   Since the $B$ coefficient goes like 
$g_1g_x\epsilon\,m_{3/2}^2\ln(\Lambda/m_{3/2})$
while $A$ goes like $g_1^2\epsilon^2m_{3/2}^2\ln(\Lambda/m_{3/2})$,
the natural size of $A$ relative to $B$ is down by a factor
of $(\epsilon/g_1g_x)\,g_1^2(Q)$.  Thus, a mild cancellation is 
usually needed between the two contributions to $B$ to keep it 
relatively small, corresponding to a fine tuning on the order
of $\epsilon$.  From this point of view, slightly larger values
of $\epsilon$ are favourable.

\subsection{Mass Sum Rules}

  The scalar superpartner masses vary widely over the $A$-$B$ plane,
and the third generation squarks can be significantly lighter than those
of the first two.  We will discuss their spectrum in more detail below.
Despite this variation, the scalar superpartner masses of the first two
generations satisfy a set of simple sum rules related to the $A$-$B$
parametrization described above.

  Only one field, $E^c$, has hypercharge such that $Y = Y^2$, so
at least three soft masses are needed to eliminate \emph{both} $A$ and $B$.
Up to an overall normalization, the complete set of soft mass combinations
that are independent of the $x$ sector is then given in
Table~\ref{tab:sumrules} including the predictions from our model;
these depend only on the standard AMSB expressions for the
soft masses~\cite{Randall:1998uk, Giudice:1998xp}.

  Note that the ordinary AMSB soft masses depend only on the three gauge
couplings and $\mgrav$.  However, the dependence on the $U(1)_x$ gauge
coupling is in all cases proportional to $Y^2$; this is the same as
the dependence on $A$, so $g_1$ will cancel by construction from
the functions we have considered.  All the masses depend on $\mgrav$
identically, and so only two of the ten combinations are linearly
independent.  This can be seen clearly in the second column
of Table ~\ref{tab:sumrules}.

  While the values of the sum rules are independent of the $x$ sector,
the ones that will be most useful in practice are not.  The spectrum 
will determine which three scalar masses can be measured first, 
and thus compared
to the predictions of Table~\ref{tab:sumrules}. The question of
the extent to which these predictions can be tested at the LHC
or other possible future experiments is deferred to future work.

\begin{table}
  \centering
  \begin{tabular}{|c|c|c|}
    \hline
    & \multicolumn{2}{c|}{Prediction} \\
    \cline{2-3}
    Scalar Mass Sum Rule & Analytic $\times \frac{\mgrav^2}{(4\pi)^4}$ & Numerical \\
    \hline
    $24 m_Q^2 + m_u^2 - 10m_d^2$ & $120 g_3^4 - 36 g_2^4$ & (4.8 TeV)$^2$ \\
    $3 m_Q^2 - 3 m_u^2 + 5 m_L^2$ & $ - 12 g_2^4 $ & - (550 GeV)$^2$ \\
    $24 m_Q^2 + 3 m_u^2 - 2 m_e^2$ & $216 g_3^4 - 36 g_2^4$ & (6.5 TeV)$^2$ \\
    $15 m_Q^2 - 6m_d^2 + m_L^2$ & $72 g_3^4 - 24 g_2^4$ & (3.7 TeV)$^2$ \\
    $24 m_Q^2 - 15 m_d^2 + m_e^2$ & $72 g_3^4 - 36 g_2^4$ & (3.7 TeV)$^2$ \\
    $27 m_Q^2 + 5 m_L^2 - 2 m_e^2$ & $216 g_3^4 - 48 g_2^4$ & (6.5 TeV)$^2$ \\
    $5 m_u^2 - 2 m_d^2 - 8 m_L^2$ & $24 g_3^4 + 12 g_2^4$ & (2.3 TeV)$^2$ \\
    $m_u^2 + 5 m_d^2 - m_e^2$ & $48 g_3^4$ & (3.1 TeV)$^2$ \\
    $27 m_u^2 - 40 m_L^2 - 2 m_e^2$ & $216 g_3^4 + 60 g_2^4$ & (6.7 TeV)$^2$ \\
    $27 m_d^2 + 8 m_L^2 - 5 m_e^2$ & $216 g_3^4 - 12 g_2^4$ & (6.4 TeV)$^2$ \\
    \hline
  \end{tabular}
  \caption{The ten combinations of squark and slepton masses that are independent of
the details of the $x$ sector, and the predictions for those values ($\mgrav =
60$~TeV).}\label{tab:sumrules}
\end{table}

\section{Phenomenology\label{sec:coll}}

  The modified AMSB scenario under consideration has gaugino masses
as in minimal AMSB together with a broad range of scalar soft masses
that can be characterized by the two parameters $A$ and $B$.
In this section we investigate the collider and cosmological
implications of the resulting superpartner spectra.
A key feature of the spectra is that the LSP is almost always
a predominantly Wino neutralino accompanied by a nearly degenerate
chargino state.  The small mass difference makes the chargino
metastable and allows it travel a macroscopic distance (typically a few cm)
before decaying.  We begin by discussing the implications of this effect
on the collider signals of this scenario.  Next, we study the mass spectrum
and the LHC collider phenomenology of several specific benchmark points 
within the $A$-$B$ plane, which we find can be significantly different 
from those considered in other modified AMSB 
scenarios~\cite{Rattazzi:1999qg,Baer:2000bs,Barr:2002ex,Baer:2010kd}. Finally, we comment on the effects of a mostly-wino LSP on cosmology.

\subsection{Charginos and Charged Tracks}
\label{sec:CMSP}

\begin{figure}[ttt]
\begin{center}
\includegraphics[width=0.47\textwidth]{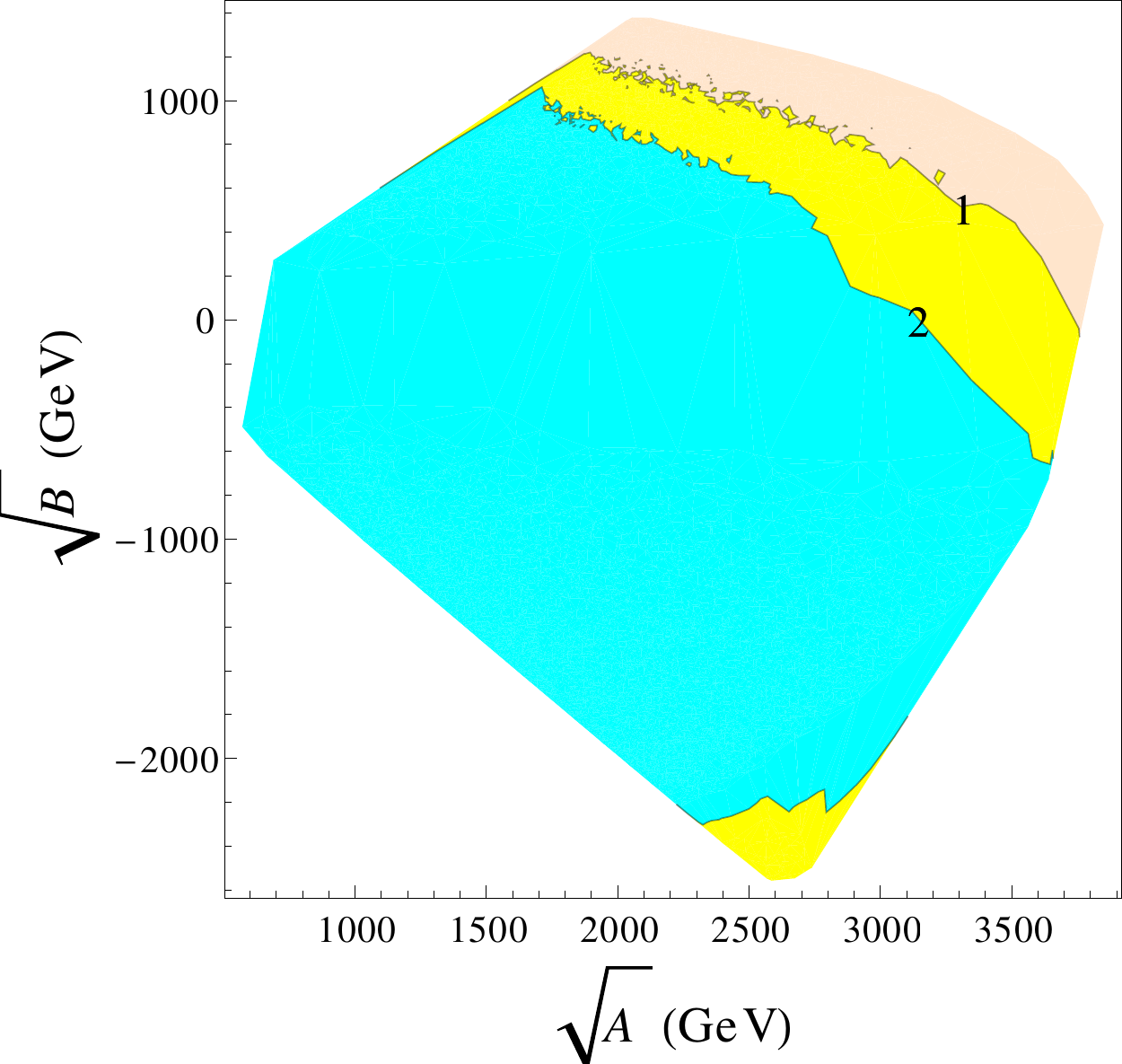}
\caption{Chargino decay lengths $c\tau$ in cm throughout
the $A$-$B$ plane for $m_{3/2}=60\,\tev$ and $\tan\beta=10$.}
\label{fig:CMSP}
\end{center}
\end{figure}

  As in minimal AMSB, the gaugino mass spectrum in our scenario has a
Wino that is considerably lighter than the other gauginos.
In most of the parameter space the $\mu$ term is also much larger 
than $M_2$.  These two features lead to a chargino
NLSP that is nearly degenerate with a mostly-Wino neutralino LSP.
If the mass splitting exceeds the charged pion mass $\Delta m > m_{\pi}$, 
the dominant decay mode is $\til \chi_{1}^{\pm} \ra \til \chi_{1}^{0} \pi^{\pm}$.
The rate for this decay is often quite slow and the chargino can
travel a macroscopic distance to produce a charged track several 
centimeters long~\cite{Gherghetta:1999sw,Gunion:2001fu,Ibe:2006de}.
If $\Delta m < m_{\pi}$, the even slower three-body mode
$\til \chi_{1}^{\pm} \ra \til \chi_{1}^{0} e \nu_{e}$
dominates and gives rise to a long charged track that extends beyond
the muon chamber~$\cite{Thomas:1998wy}$.   Such long-lived charginos
could be seen in standard searches for metastable charged particles.
On the other hand, short charged track stubs from the two-body decay
require a different set of triggers, as discussed in Ref.~\cite{Buckley:2009kv}.

  Evidently the chargino-neutralino mass difference plays a crucial
role in determining the collider phenomenology of this and other AMSB-like
constructions.  Relative to minimal AMSB, the value of $\mu$ (as determined
by electroweak symmetry breaking and the predicted scalar soft masses)
and the masses of the scalar superpartners can be considerably different
in our scenario, and this can modify the chargino-neutralino mass difference.
In most of the $A$-$B$ plane we find that $\mu$ is very large relative
to $M_2$ leading to a very small tree-level mass splitting $\Delta m$,
typically well below the pion mass.  However, as discussed
in Ref.~\cite{Gherghetta:1999sw}, finite loop corrections involving the 
scalar superpartners tend to increase the mass
splitting above the pion threshold~\cite{Pierce:1996zz}.

  We find that one-loop corrections to the MSSM spectrum
push $\Delta m > m_{\pi}$ throughout the entire $A$-$B$ plane
for any reasonable values of $m_{3/2}$ and 
$\tan\beta$~\cite{Gherghetta:1999sw}.
This leads to dominant two-body chargino decays with decay 
lengths $c\tau$ in the range of a few centimeters,
as illustrated in Fig.~\ref{fig:CMSP} for $m_{3/2}=60\,\tev$
and $\tan\beta=10$.  
The mass splitting is largest towards the upper-right section 
of this plane due to smaller values of $\mu$ that occur there. Additionally,
light mostly left-handed stops and sbottoms increase the mass splitting
in the bottom-right section of the plane.

  Charged track stubs of this length could be observed in the inner 
detectors of ATLAS and CMS, albeit with some difficulty.  Without a 
dedicated search for them, the light charginos contribute to missing energy 
in the event since the pion emitted in their decay is very soft.
As such, the standard SUSY search techniques can be applied to this
scenario as well, and we will treat both states as contributing to
missing energy for the remainder of this discussion.  
However, the track stubs from chargino decays could potentially be 
extracted from the inner detector if a specific search is made
for them, provided they are accompanied by a hard jet or photon 
to serve as a trigger~\cite{Gherghetta:1999sw,
Baer:2000bs,Barr:2002ex,Chen:1995yu}.

\subsection{Benchmark Points}

\begin{figure}[ttt]
\begin{center}
\includegraphics[width=0.43\textwidth]{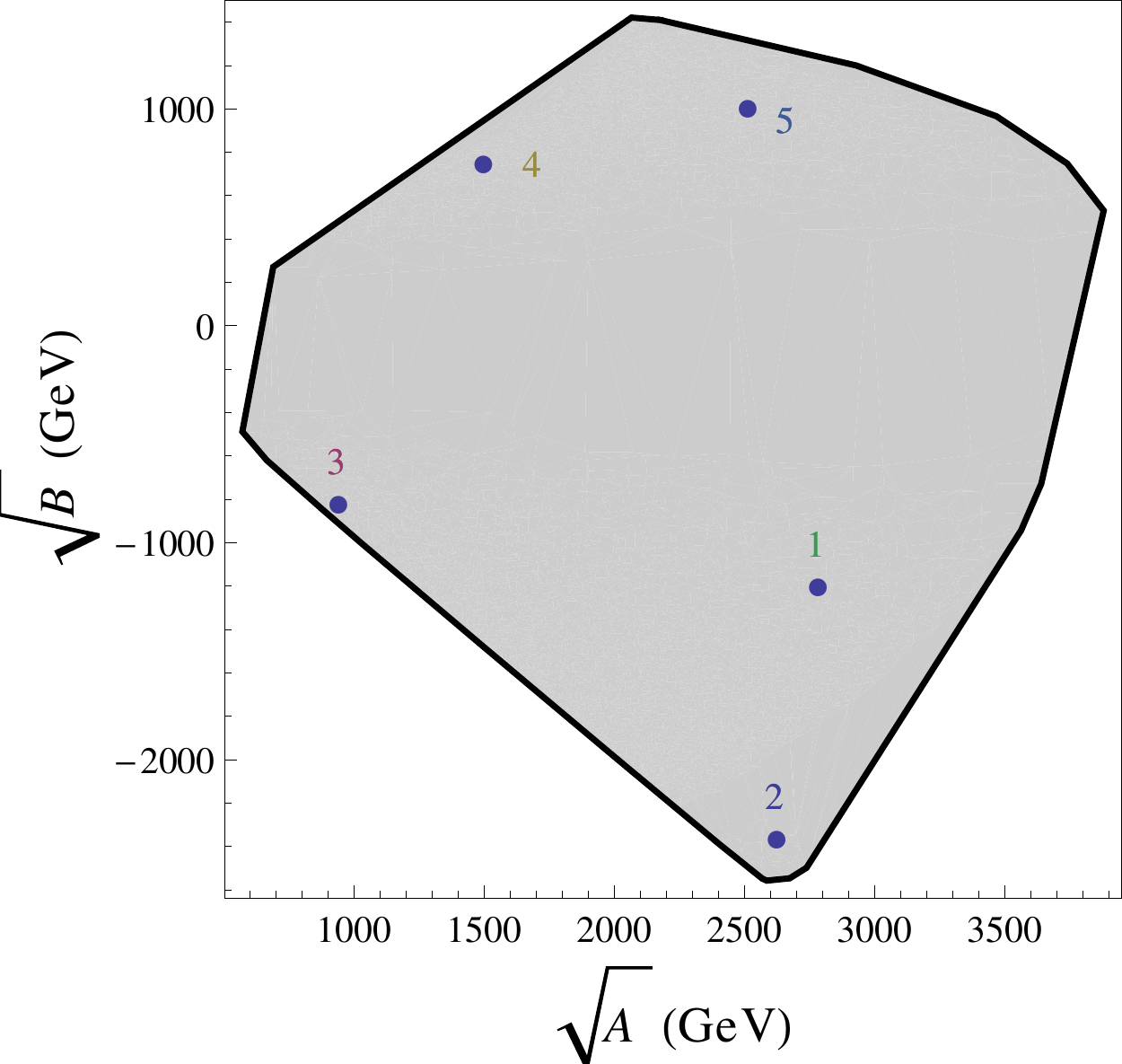}
\hspace{2cm}
\includegraphics[width=0.4\textwidth,viewport={130 300 600 780}]{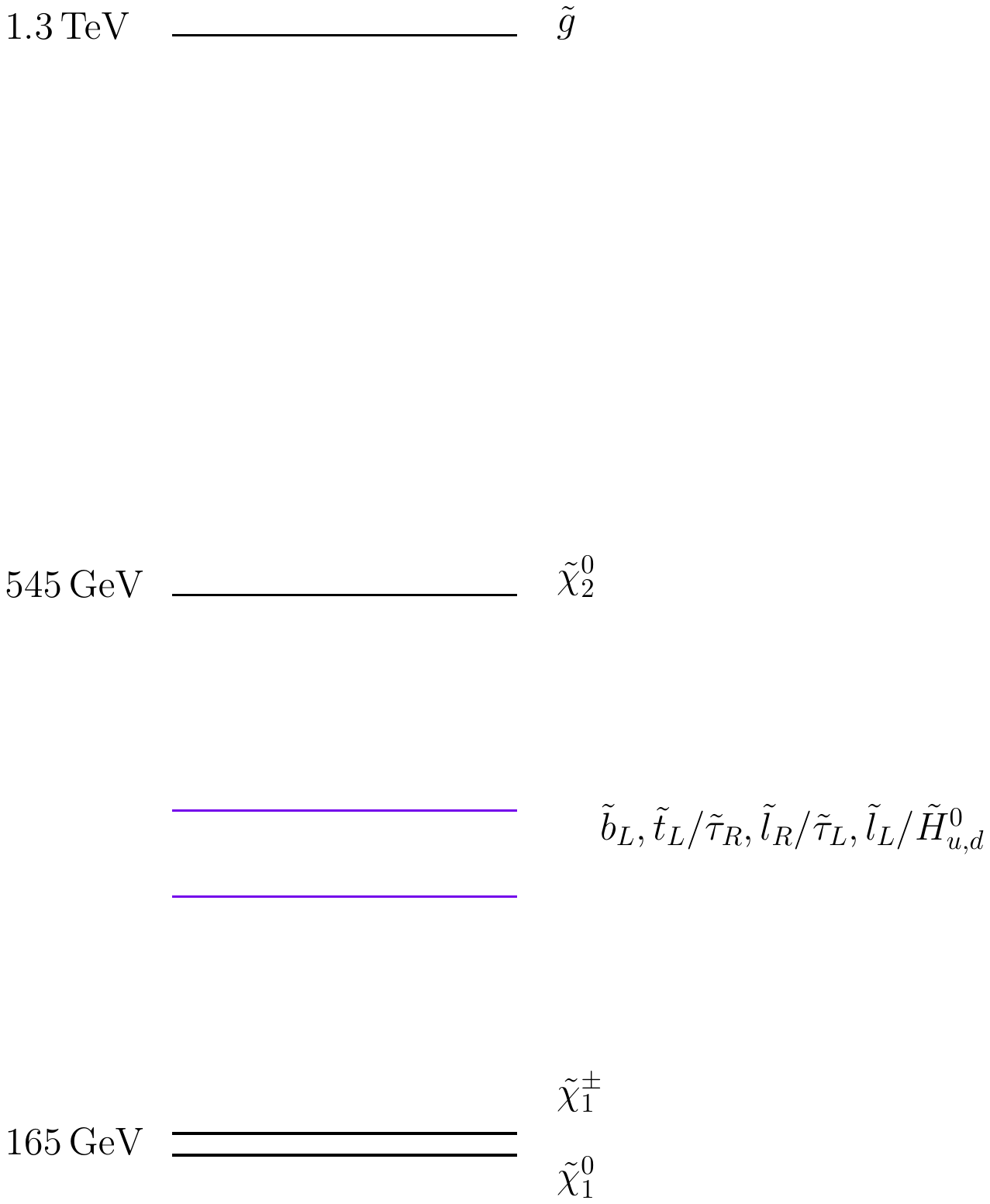}
\caption{Locations of the five benchmark points in the $A$-$B$ plane
with $m_{3/2}=60\,\tev$ (left), and the schematic particle spectrum
of these points (right).}
\label{fig:Benchpts}
\end{center}
\end{figure}

  We study five benchmark points that cover a broad range of the
qualitative collider signatures that can arise in this scenario.
Recall from Fig.~\ref{fig:lightestsqkslep} that the left-handed squark
soft masses increase from lower right to upper left in the $A$-$B$ plane,
the right-handed slepton soft masses increase from upper left to lower right,
and the left-handed slepton soft masses increase from lower left to
upper right.  Our benchmark points are chosen along these three gradients,
and we show their locations in the $A$-$B$ plane in the left
panel of Fig.~\ref{fig:Benchpts}.  For all these points, we choose
$m_{3/2} = 60\,\tev$ and $\tan\beta = 10$, and we label them as
XAMSB$_i$ with $i=1\!-\!5$.  Our choice of $m_{3/2}$ provides a
superpartner spectrum that is accessible at the LHC while being
consistent with existing LHC searches.

  For each benchmark point we generate decay tables using
SUSY-HIT~$\cite{Djouadi:2006bz}$.  The leading-order LHC7 
production cross sections ($pp$ collider at $\sqrt{s} = 7\,\tev$) 
are computed with Pythia~6.4~$\cite{Sjostrand:2006za}$.
We also cross-check our results with BRIDGE $\cite{Meade:2007js}$
and MadGraph $\cite{Alwall:2007st}$, and find consistent behaviour.
The generic features of the spectra for all these points are illustrated
schematically in the right panel of Fig.~\ref{fig:Benchpts}.
The mass spectra and dominant branching fractions are shown in
Figs.~\ref{fig:XAMSB12}-\ref{fig:XAMSB67}, while the leading superpartner
production cross sections at the LHC are listed in
Tables~\ref{tab:XAMSBtab12}, \ref{tab:XAMSBtab345}, \ref{tab:XAMSBtab67}.
In Figs.~\ref{fig:XAMSB12}-\ref{fig:XAMSB67}, the notation for branching
ratios to NLSP/LSP is in the order of heavier/lighter
SUSY final state. 

  All these points have similar gaugino spectra, with a mostly-Wino
LSP and chargino NLSP with masses close to 165~GeV, a mostly-Bino
neutralino with mass near 545~GeV, and a gluino with mass near 1.3~TeV.
On the other hand, the masses of Higgsino-like states as well as the 
scalar spectra vary widely.  We will discuss the details below.

\subsubsection{$XAMSB_{1}$  $(\sqrt{A} = 2775\,\gev$, $sgn \sqrt{B}= -1203\,\gev$) }

\begin{figure}[ttt]
\begin{center}
\includegraphics[width=0.47\textwidth,viewport={55 250 585 765}]{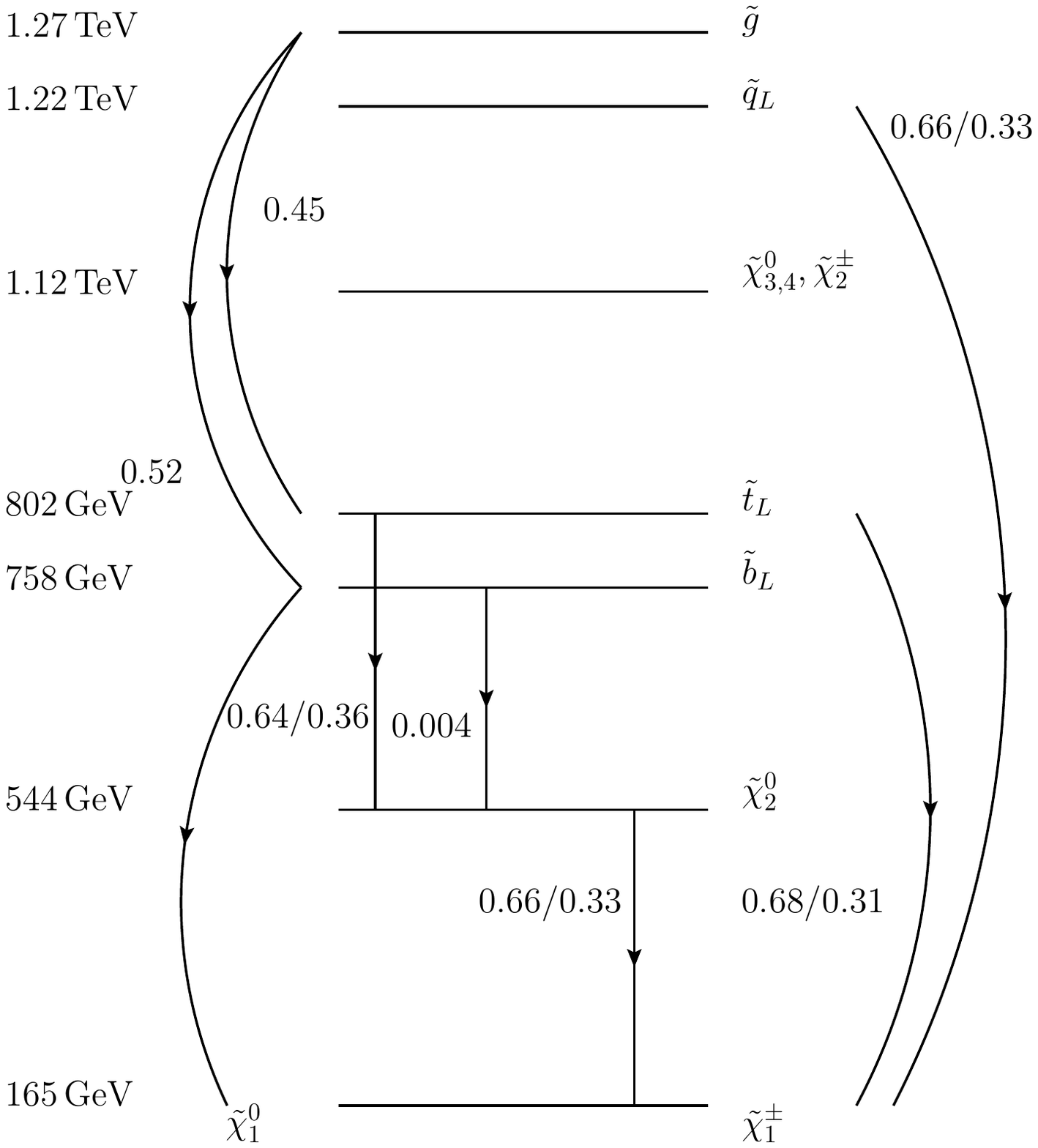}
\hspace{0.7cm}
\includegraphics[width=0.47\textwidth,viewport=
{55 265 585 765}
]{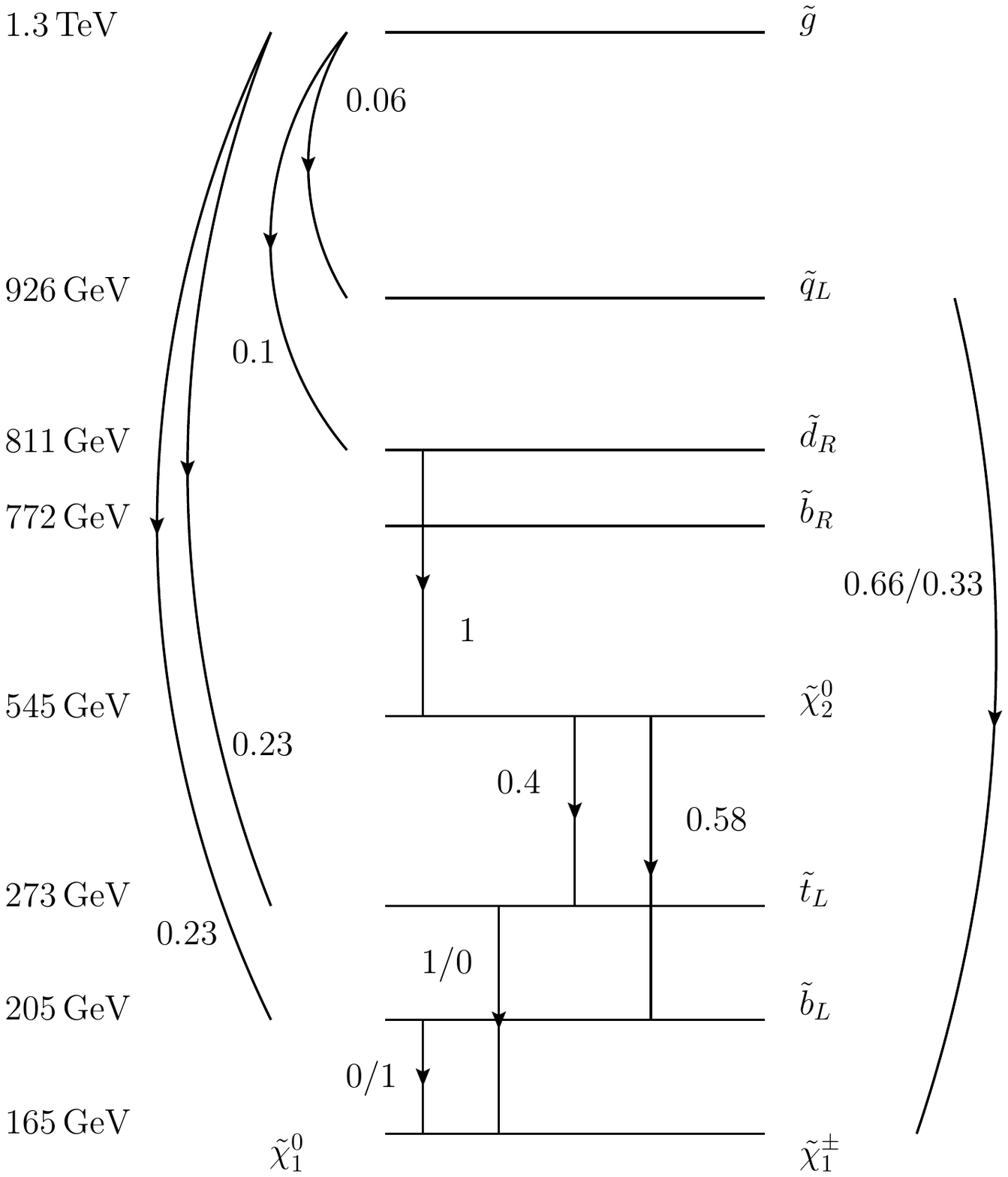}
\caption{Mass spectrum and decay branching fractions for the benchmark
points $XAMSB_{1}$ (left) and $XAMSB_2$ (right).}
\label{fig:XAMSB12}
\end{center}
\end{figure}

The  $XAMSB_{1}$ benchmark point has a mostly left-handed sbottom
with a mass near 750~GeV and a slightly heavier left-handed stop 
near $800\,\gev$ as the lightest squarks.  The other squarks
and the gluino have masses in excess of $1.2\,\tev$,
as seen in the left panel of Fig.~\ref{fig:XAMSB12}.

\begin{table}[ttt]
  \centering
  \begin{minipage}{0.45\textwidth}
    \centering
    \begin{tabular}{|c|c|}
      \hline
      $\sigma_{\rm SUSY}$ & 1.35 pb \\
      \hline
      $\til\chi_{1}^{0}\til\chi_{1}^{\pm} + \til\chi_{1}^{\mp}\til\chi_{1}^{\pm}$ & 1.34 pb \\
      $\til q_{L} \til q_{L}^{(*)}$ & 2.4 fb\\
      $\til\chi_{1}^{0}\til q_{L} + \til\chi_{1}^{\pm}\til q_{L}$ & 1.36 fb \\
      $\til g \til q_{L}$ & 1.2 fb \\
      $\til g \til g$ & 0.1 fb \\
      $\til b_L\til b_L^{(*)}$& 0.9 fb \\
      $\til t_L \til t_L^{(*)}$& 0.7 fb\\
      \hline
    \end{tabular}
  \end{minipage}
  \begin{minipage}{0.45\textwidth}
    \centering
    \begin{tabular}{|c|c|}
      \hline
       $\sigma_{\rm SUSY}$ & 8.55 pb \\
      \hline
      $\til\chi_{1}^{0}\til\chi_{1}^{\pm} + \til\chi_{1}^{\mp}\til\chi_{1}^{\pm}$ & 1.25 pb \\
      $\sbsb$ & 6.0 pb \\
      $\stst$ & 1.2 pb \\
      $\til q_{L} \til\chi_{1}^{0}$ & 0.09 pb\\
      $\til q_{L} \til q_{L}^{(*)}$ & 0.02 pb\\
      $\til q_{L} \bar{\til q}_{R}$ & 7.0 fb \\
      $\til d_{R} \til d_{R}^{(*)}$ & 0.3 fb\\
      \hline
    \end{tabular}
  \end{minipage}
  \caption{Production cross sections for $XAMSB_{1}$ and $XAMSB_{2}$.}
  \label{tab:XAMSBtab12}
\end{table}

  The production of superpartners is dominated by
electroweak chargino and neutralino pair creation processes,
and is given in Table~\ref{tab:XAMSBtab12}.  
The lightest neutralino will be invisible, as will the 
lightest chargino up to an accompanying short charged track stub 
and a very soft pion.  To observe these electroweak production processes, 
an additional radiated hard jet or photon would be 
required~\cite{Baer:2000bs,Barr:2002ex,Chen:1995yu}.  
The associated squark/-ino production modes would yield a similar
monojet signal, but have very small cross sections 
(given in Table~\ref{tab:XAMSBtab12}).

  The rates for QCD production of squarks and gluinos are suppressed
by the heavier masses of these states.  For the gluino and the light-flavour
squarks, the production rates are safely below the current limits
from LHC searches for jets and missing 
energy~(MET)~\cite{atlas-conf-2011-086,cms-pas-sus-11-003,cms-pas-sus-11-004,cms-pas-sus-11-005}. 
This is also true for the lighter stop and sbottom states,
which have lower production rates for a given mass since
they are not created efficiently via $qq$ initial states.
In particular, the sbottom and stop states are heavy enough to avoid the
LHC limits from searches for $b$ jets and MET~\cite{atlas-conf-2011-098} 
as well as inclusive searches for jets and MET.

\subsubsection{$XAMSB_2$ ($\sqrt{A}=2617\,\gev$, $sgn\sqrt{B}=-2364\,\gev$)}

 The spectrum for $XAMSB_{2}$ has a mostly left-handed sbottom 
and stop that are light relative to the other squarks.  As shown in 
Table~\ref{tab:XAMSBtab12}, this leads to very large production 
cross-sections.  The smaller mass differences between 
these squarks and the Wino-like chargino 
and neutralino states cause them to decay primarily into $b$ quarks 
and $\tilde{\chi}_1^0$ (for $\tilde b_1$) or $\tilde{\chi}_1^+$ 
(for $\tilde{t}_1$) rather than the kinematically-suppressed top modes. 
As such, we expect that this point can be probed by LHC searches
in the inclusive jets plus MET 
channels~\cite{atlas-conf-2011-086,cms-pas-sus-11-003,cms-pas-sus-11-004,cms-pas-sus-11-005},
in addition to searches for bottom jets and MET~\cite{atlas-conf-2011-098}.
However, despite the large production rates the relatively small
mass differences between the stop/sbottom and the chargino/neutralino
make these squarks difficult to detect.  For this reason, these states
are not constrained by Tevatron searches~\cite{Abazov:2010wq,
Aaltonen:2010dy}, and we expect the same to hold for existing LHC searches.

  The $\tilde{d}_R$ state is also not overly heavy at the $XAMSB_2$
sample point.  Its production is too small to be observed with existing
analyses.  Even so, with future searches in mind we note that it decays 
through an extended cascade, going first to $d+\til\chi_2^0$, 
followed by $\til\chi_2^0$ decaying to $\tilde{t}_L+t$
or $\tilde{b}_L+b$, which then go to $b+(N)LSP$.  The net final
state therefore consists of multiple light and $b$ jets, possibly
some leptons, and missing energy.

\subsubsection{$XAMSB_{3}$
and $XAMSB_{4}$\\  
${}$\hspace{-0.8cm}~($\sqrt{A}=938\,\gev$, $sgn\sqrt{B}=-809\,\gev$)  and 
($\sqrt{A}=1491\,\gev$, $sgn\sqrt{B}=749\,\gev$)}

\begin{figure}[ttt]
\begin{center}
\includegraphics[width=0.47\textwidth,viewport={65 200 570 770}]{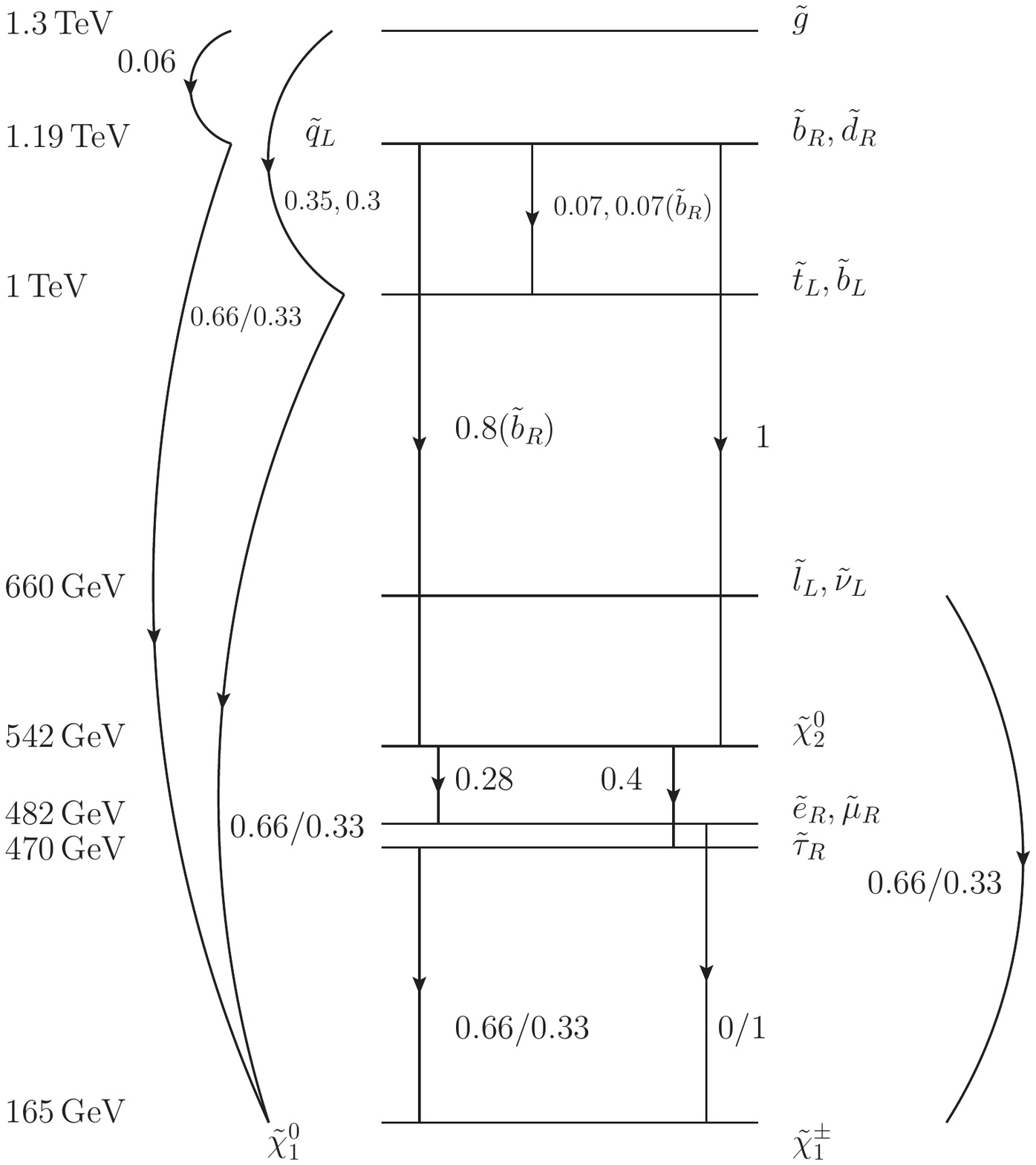}
\includegraphics[width=0.47\textwidth,viewport={86 220 578 690}]{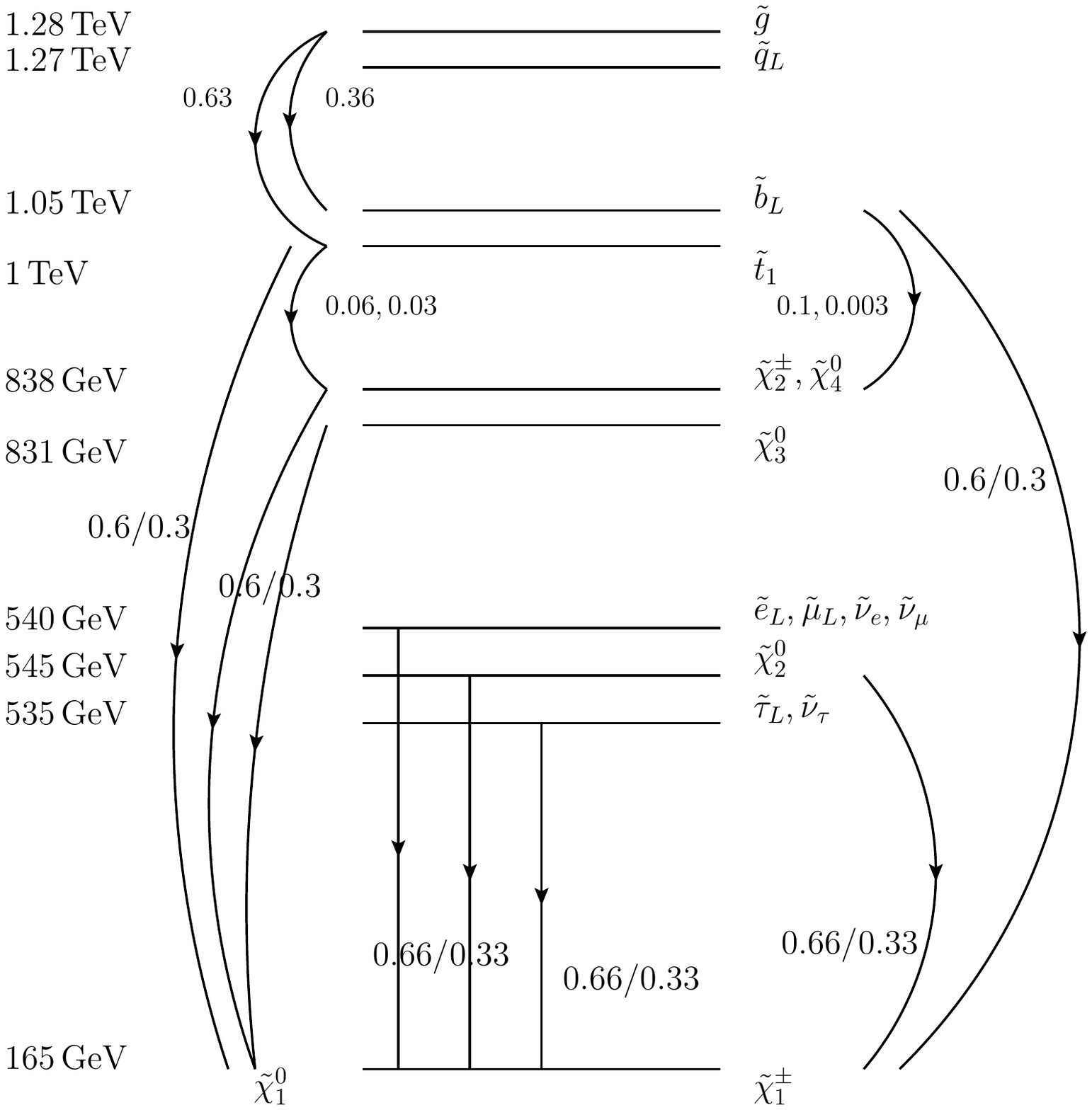}
\caption{Spectrum and branching fractions for $XAMSB_{3}$ and $XAMSB_{4}$.}
\label{fig:XAMSB45}
\end{center}
\end{figure}

\begin{table}[ttt]
  \centering
  \begin{minipage}{0.45\textwidth}
    \centering
    \begin{tabular}{|c|c|}
      \hline
       $\sigma_{\rm SUSY}$ & 1.35 pb \\
      \hline
      $\til\chi_{1}^{0}\til\chi_{1}^{\pm} + \til\chi_{1}^{\mp}\til\chi_{1}^{\pm}$ & 1.33 pb\\
      $\til q_{L} \til q_{L}$ & 2.8 fb\\
      $\til q_{L} \til \chi_{1}^{\pm}$ & 1.5 fb\\
	$\til q_{R} \til q_{R} $ & 0.7 fb\\
      \hline
    \end{tabular}
  \end{minipage}
  \vspace{1cm}
  \begin{minipage}{0.45\textwidth}
    \centering
    \begin{tabular}{|c|c|}
      \hline
      $\sigma_{\rm SUSY}$ & 1.4 pb \\
      \hline
      $\til\chi_{1}^{0}\til\chi_{1}^{\pm} + \til\chi_{1}^{\mp}\til\chi_{1}^{\pm}$ & 1.39 pb\\
      $\til q_{L} \til q_{L}$ & 1.6 fb\\      
      $\til q_{R} \til q_{R}$ & 0.35 fb\\
      $\til q_{R} \til g$ & 0.3 fb\\
      \hline
    \end{tabular}
  \end{minipage}
  \caption{SUSY production cross section for
$XAMSB_{3}$ and $XAMSB_{4}$}
  \label{tab:XAMSBtab345}
\end{table}

  Points $XAMSB_3$ and $XAMSB_4$ are similar in that they both have
relatively light sleptons, lighter than the mostly-Bino sate, 
as can be seen in Fig.~\ref{fig:XAMSB45}.
In the case of $XAMSB_3$ the lighter sleptons are mainly right-handed,
while for $XAMSB_4$ they are predominantly left-handed.  The presence
of such light sleptons offers the possibility of additional leptons
in cascade decays.  Both points also have all squarks heavier than about
1 TeV, and are therefore expected to be consistent with existing LHC
searches for jets plus MET.

  The mass spectrum of $XAMSB_3$ leads to final states with multiple
leptons.  The right-handed squarks decay preferentially to the
Bino-like $\tilde{\chi}_2^0$ neutralino rather than the lighter
Wino-like neutralinos and charginos.  Subsequently, $\tilde{\chi}_2^0$
decays most of the time to the lighter charged sleptons by emitting
the corresponding charged lepton.  Yet another charged lepton (of opposite sign)
will be produced if the slepton decays to the neutralino LSP.  
The dilepton invariant mass in this case will provide an efficient
probe of the underlying slepton masses.

  In contrast, the relatively light (mostly left-handed) sleptons
of $XAMSB_4$ are not populated very efficiently by cascade decays.
The $\tilde{\chi}_2^0$ mode decays primarily via a $W^{\pm}$ or $Z^0$
in a 2:1 ratio to $\til\chi_1^{\pm}$ or $\til\chi_1^0$, and are therefore
a less efficient source of leptons.  The most promising search
strategy for this parameter point is thus likely to be jets
plus MET.

\subsubsection{$XAMSB_{5}$ ($\sqrt{A}=2506\,\gev$, $sgn\sqrt{B}=1004\,\gev$)}

\begin{figure}[ttt]
\begin{center}
\includegraphics[width=0.45\textwidth,viewport={70 290 520 765}]{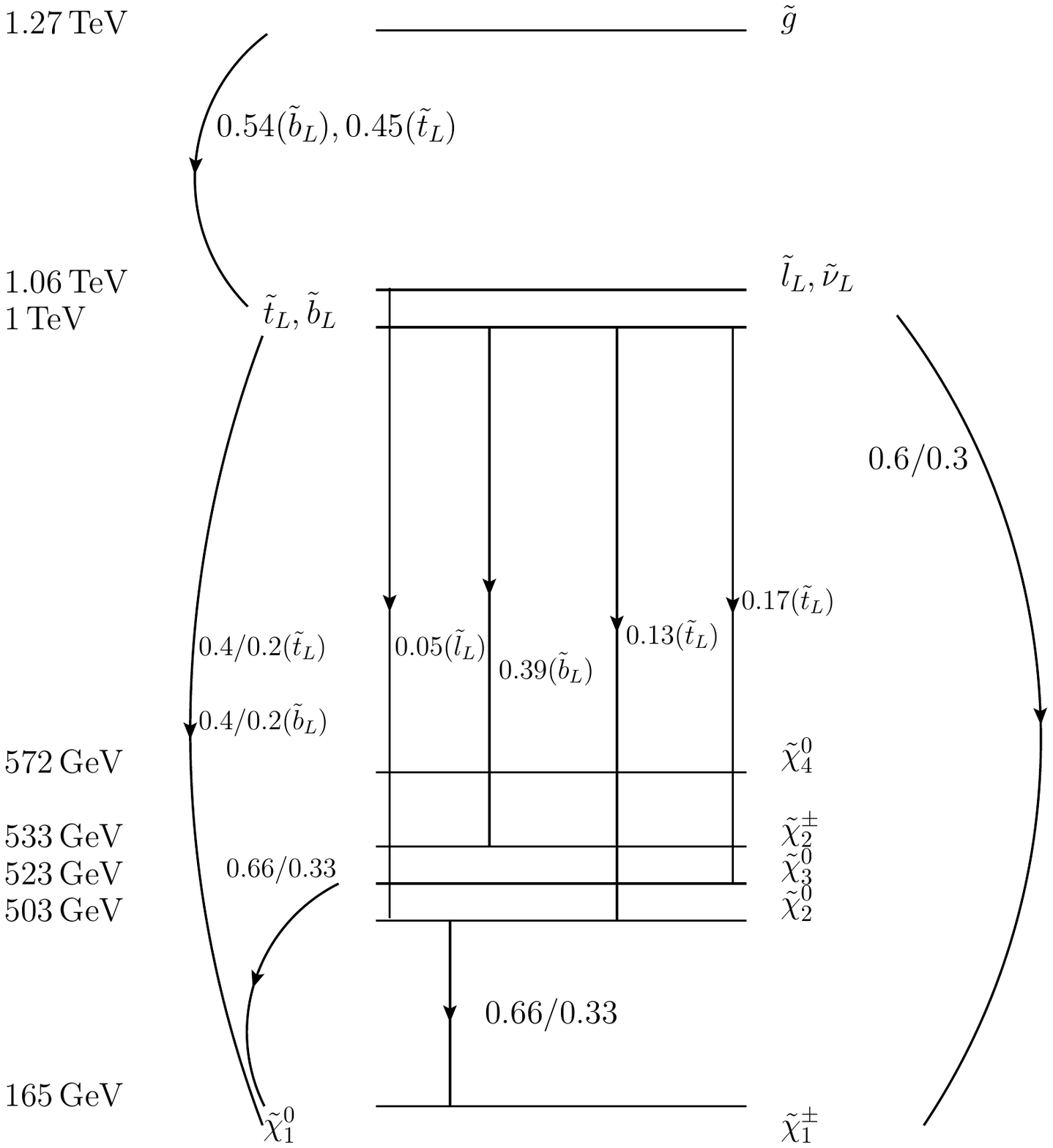}
\caption{Spectrum and branchings for $XAMSB_{5}$.}
\label{fig:XAMSB67}
\end{center}
\end{figure}

  Relative to the other sample points, the distinctive feature of 
$XAMSB_{5}$ is that it has lighter higgsinos.  As shown in 
Fig.~\ref{fig:XAMSB67}, these give rise to multiple chargino and 
neutralino states with masses close to 600 GeV, well below the 
masses of all the scalars.  In comparison, the lightest squarks 
are the mostly-left-handed stop and sbottom, with masses close to 1 TeV.

  The lighter stop and sbottom states have significant
branching fractions to decay to Higgsino-like charginos 
or neutralinos, on account of the large top Yukawa coupling.
Their decays will involve a top or bottom quark.
Higgsino-like charginos and neutralinos will subsequently
decay to the lightest Wino-like states by emitting a 
$W$, a $Z$, or the lightest Higgs boson $h^0$. 
Cascades for this point are therefore expected 
to include multiple bottom quarks, both from direct 
bottoms, top decays, and Higgs decays.

\begin{table}
  \centering
  \begin{minipage}{0.45\textwidth}
    \centering
    \begin{tabular}{|c|c|}
      \hline
       $\sigma_{\rm SUSY}$ & 1.5 pb \\
      \hline
      $\til\chi_{1}^{0}\til\chi_{1}^{\pm} + \til\chi_{1}^{\mp}\til\chi_{1}^{\pm}$ & 1.49 pb\\
      $\til\chi_{3}^{0} \til\chi_{2}^{\pm}$ & 1.4 fb\\
      $\til\chi_{2}^{0} \til\chi_{2}^{\pm}$ & 0.9 fb\\
	$\sbsb$ & 0.04 fb\\
      $\stst$ & 0.03 fb\\
      \hline
    \end{tabular}
  \end{minipage}
  \caption{SUSY production cross sections for $XAMSB_{5}$}
  \label{tab:XAMSBtab67}
\end{table}

\subsection{Comments on Dark Matter}

  The cosmological behaviour of this extended AMSB scenario is expected
to be very similar to many other supersymmetric theories with a gaugino
mass spectrum following the minimal AMSB scenario.  The obvious dark
matter candidate is the mostly-Wino neutralino LSP.
This state annihilates (and coannihilates with the nearly degenerate
charginos) very efficiently in the early Universe.
The resulting dark matter relic density is smaller than the observed
value for $M_2$ below about $2.5\,\tev$~\cite{ArkaniHamed:2006mb}.\footnote{
Note as well that the metastable charginos decay relatively quickly
on cosmological time scales, and certainly well before the onset of
primordial nucleosynthesis.}  For $M_2 \simeq 2.5\,\tev$ an acceptable
relic density is obtained, but the entire superpartner mass spectrum
in this case is very heavy relative to the electroweak scale.

  An acceptable mostly-Wino neutralino relic density can arise
for smaller neutralino masses if these states are produced non-thermally,
such as through the decay of a moduli
field~\cite{Moroi:1999zb,Acharya:2009zt}.  Due to the relatively large
annihilation cross section of such states, they have been put forward~\cite{
Grajek:2008pg,Kane:2009if} as a possible explanation for the apparent
excess of positrons seen by the PAMELA experiment~\cite{Adriani:2008zr}.
However, this proposal is constrained by the apparent absence of an
excess in antiproton cosmic rays in the PAMELA data~\cite{Adriani:2010rc},
by searches by the Fermi Space Telescope for gamma rays
that are monoenergetic~\cite{Abdo:2010nc} or originating in dwarf
spheroidal galaxies~\cite{Abdo:2010ex},
and from limits on energy injection at recombination based
on measurements of the cosmic microwave background~\cite{Slatyer:2009yq}.
Mostly-Wino DM can also be consistent with existing bounds from
direct detection (such as XENON~100~\cite{Aprile:2011hi}) provided 
$|M_2/\mu| \ll 1$~\cite{Moroi:1999zb,Baer:2010kd}.

\section{Conclusion\label{sec:conc}}

  In this work we have proposed and investigated a simple
extension of minimal AMSB that can give rise to a viable
spectrum of MSSM superpartners while not introducing dangerous
new sources of flavour mixing.  A sequestered MSSM sector couples
via a gauge kinetic mixing coupling of hypercharge to a new,
unsequestered $U(1)_x$ gauge multiplet.  

  We find that the new gauge multiplet provides an additional source 
of supersymmetry breaking that can render all MSSM squark and slepton
soft masses positive and produce a phenomenologically viable spectrum.
The gaugino masses in this scenario are nearly identical to those in
minimal AMSB, while the scalar masses cover a diverse range of values.

  The LSP in this scenario is a mostly-Wino neutralino that is nearly
degenerate in mass with a mostly-Wino chargino.  This chargino can 
travel a macroscopic distance of a few centimeters after being produced
in a high-energy collider, giving rise to charged track stubs.
The scalar spectrum in this scenario tends to yield top- and bottom-rich
cascade decays at the LHC together with missing energy.  Multi-lepton
states can arise in certain cases as well.  Measuring the masses of
a variety of these states could help to identify this scenario using
LHC data as the soft scalar masses satisfy a number of simple sum rules.

\section*{Acknowledgements}

  We thank Thomas Gr\'{e}goire, John Ng, Robert McPherson, David Poland,
Yael Shadmi, and Scott Watson for helpful discussions.  
DM would also like to thank 
the CERN theory group for their hospitality while this work was being 
completed.  This research was supported by the National Sciences and 
Engineering Research Council of Canada~(NSERC).

\appendix

\section{Soft Masses and RG Running\label{sec:apprg}}

  The effects of the $x$ sector on the RG running of the MSSM soft
masses can be derived in a simple way by using analytic continuation
methods in superspace~\cite{Giudice:1997ni,ArkaniHamed:1998kj}.
For this, it is convenient to describe the approximately globally
supersymmetric theory valid below $\Lambda$ with a renormalized
1PI effective Lagrangian of the form
\bea
\mathscr{L} &=&
\int d^4\theta\left(
Z_i\phi_i^{\dagger}e^{2Y_iV_Y}\phi_i
+ Z_pH_p^{\dagger}e^{2x_pV_x}H_p
\right) + \left(\int d^2\theta\;W + h.c.\right)
\label{hollag}\\
&&+\int d^4\theta\left(\xi_YV_Y+\xi_xV_x\right)
+ \int d^2\theta\left(\frac{1}{4G_Y^2}B^{\alpha}B_{\alpha}
+ \frac{1}{4G_x^2}X^{\alpha}X_{\alpha}
+ \frac{\epsilon_h}{2}X^{\alpha}B_{\alpha}\right) + (h.c.) \ .
\nnmb
\eea
Soft supersymmetry breaking is contained within non-vanishing auxiliary
components of the renormalized couplings, which are now elevated
to real ($Z_i$) and chiral ($G_i$, $\epsilon_h$) superfields.
The $\theta^4$ component of $\ln Z_i$  contributes to
the soft scalar squared mass of $\phi_i$,
\beq
\Delta m_i^2 = -\ln Z_i|_{\theta^4} \ ,
\label{zmsq}
\eeq
while a trilinear $A$-term of the form $\phi_i\phi_j\phi_k$ comes from
\beq
A_{ijk} = \left(\ln Z_i+\ln Z_j+\ln Z_k\right)|_{\theta^2} \ .
\eeq
In the gauge sector,
we identify the physical gauge coupling and gaugino mass with
a real superfield $R$ given by
\beq
R_a \equiv \frac{1}{2G_a^2}+\frac{1}{2G^{2\dagger}_a}
-\sum_i\frac{q_i^2}{8\pi^2}\ln Z_i + \ldots \ ,
\eeq
with the sum running over all fields with charge $q_i$ under the $a$-th
gauge group and the omitted terms are of higher order in the gauge coupling.
From this vector superfield, we obtain the physical gauge coupling
and gaugino mass according to
\beq
R_a|_0 = \frac{1}{g_a^2},~~~~~~~ R_a|_{\theta^2} = -\frac{M_a}{g_a^2}  \ .
\eeq
The $\theta^4$ component of $R_a$ is smaller by 
a loop factor~\cite{ArkaniHamed:1998kj}.

  In the same way, we can form a real superfield $E$ corresponding
to $\epsilon_h$:
\beq
E = \frac{\epsilon_h+\epsilon_h^{\dagger}}{2}
- \sum_i\frac{q_{ix}q_{iY}}{8\pi^2}\ln Z_i + \ldots
\eeq
With this prescription, we identify the physical kinetic
mixing parameter $\epsilon$ with
\beq
\epsilon = E/\sqrt{R_xR_Y}|_0  \simeq g_xg_Y\,\text{Re}(\epsilon_h) \ .
\eeq

  The holomorphic basis of Eq.~\eqref{hollag} is useful because
it separates the running of hypercharge and $U(1)_x$.  Holomorphy
implies that the running of $G_x$, $G_Y$, and $\epsilon_h$ is
one-loop exact.  For the gauge couplings,
\beq
\frac{d}{dt}\lrf{1}{G_a^2} = \frac{b_a}{8\pi^2},~~~~~b_a= -tr(q_i^2)
= -2x_H^2,~-33/5 \ .
\label{gaugerun}
\eeq
Taking the $\theta^2$ component of Eq.~\eqref{gaugerun}
then gives the standard MSSM one-loop RG equation for $M_1$ as well as
\beq
(4\pi)^2\frac{dM_x}{dt} = 4x_H^2g_x^2M_x \ .
\eeq
Since there are no fields in the low-energy theory charged under both
hypercharge and $U(1)_x$, the holomorphic mixing $\epsilon_h$ does not
run at all~\cite{Morrissey:2009ur}.   As long as the source of 
kinetic mixing is sequestered from supersymmetry breaking, $\epsilon_h$ 
has a vanishing $\theta^2$ component and there is no mixed gaugino soft
mass in this basis.  We also have to one-loop order
\beq
\frac{d\epsilon}{dt} \simeq
\left(\frac{\beta_x}{g_x}+\frac{\beta_Y}{g_Y}\right)\epsilon
\simeq \frac{1}{4\pi^2}\left(2x_H^2g_x^2+\frac{33}{5}g_1^2\right)\epsilon \ ,
\eeq
with $\beta_a = dg_a/dt$.

  Running of the soft scalar squared masses and the trilinear $A$ terms
can be derived from the higher components of the wavefunction factors.
At one-loop order, the running of the wavefunction of the $i$-th MSSM
field is~\cite{Morrissey:2009ur}
\beq
(4\pi^2)\frac{d\ln Z_i}{dt} = (4\pi^2)\left(\frac{d\ln Z_i}{dt}\right)_{\!MSSM}
\!\!+~4Y_i^2R_Y^{-1}\left(\frac{1}{1-E^2R_x^{-1}R_Y^{-1}} - 1\right) \ ,
\eeq
where the $E^2$-dependent correction comes from resumming
double insertions of the kinetic mixing on diagrams of the form
of Fig.~\ref{fig:Mxloop}.  Expanding out the correction to quadratic
order in $\epsilon$ gives
\bea
\left.(\ldots)\right|_{\theta^2} &=&
4g_Y^2Y_i^2{\epsilon}^2(M_x+2M_1) \ ,
\\
\left.(\ldots)\right|_{\theta^4} &=&
8g_Y^2Y_i^2{\epsilon}^2\left(
|M_x+M_1|^2+2|M_1|^2
\right) \ .
\eea
Keeping only those terms that are enhanced by $|M_x|\gg |M_1|$
leads to the $M_x$-dependent parts of Eqs.~(\ref{misoft},\ref{aisoft}).

  The Lagrangian of Eq.~\eqref{hollag} also contains Fayet-Iliopoulos~(FI)
terms for $U(1)_x$ and hypercharge.
At one-loop they run according to~\cite{Jack:2000nm}
\beq
\frac{d\xi_a}{dt} = \frac{2}{(4\pi)^2}\,tr(q_a\Delta m^2) \ ,
\label{firun}
\eeq
where $\Delta m_i^2$ is the contribution to the $i$-th scalar soft
mass from the corresponding wavefunction factor, as in Eq.~\eqref{zmsq}.
For a single $U(1)$ factor, when the auxiliary $D$-fields are integrated out,
the FI terms effectively contribute to the scalar soft masses, producing
a total value of
\beq
m_i^2 = \Delta m_i^2 + q_{i,a}g_a^2\xi_a \ .
\label{mtot}
\eeq
The RG running of these two contributions can be dealt with separately,
or together in a single RG equation for $m_i^2$.  In the latter case,
the contributions of the FI term to the running are closely related to the
``S'' terms proportional to $tr(Ym^2)$ that appear in the usual RG
equations for the MSSM soft scalar masses~\cite{Martin:1997ns}.
With several $U(1)$ factors and kinetic mixing, there are additional
effects that we will describe below.

  After running from the high scale $\Lambda$ in the MSSM$\times U(1)_x$
theory down to scale $m_{3/2}$, we integrate out the $x$ sector
and continue to run the remaining soft masses in the MSSM alone.
For this, it is convenient to convert Eq.~\eqref{hollag} to a canonical
basis for all fields, as in Eq.~\eqref{eq:kinmix}.
Kinetic mixing among the gauge fields and the gauginos can
be eliminated by shifting the (now canonically normalized aside from the mixing)
vector multiplets according to
\bea
V_Y &\to& V_Y - s_{\epsilon}V_x,
\label{shift}\\
V_x &\to& c_{\epsilon}V_x,
\nnmb
\eea
where $c_{\epsilon} = 1/\sqrt{1-\epsilon^2}$ and
$s_{\epsilon}= \epsilon\,c_{\epsilon}$.
After making this shift, the vector multiplets are canonically normalized
with no gauge kinetic mixing, the fields initially charged under $U(1)_x$
now couple to the $U(1)_x$ multiplet with strength
\beq
g_xc_{\epsilon}x_p \equiv \tilde{g}_xx_p \ .
\eeq
In particular, the mass of the $U(1)_x$ vector boson is 
$2\tilde{g}_x^2x_H^2\eta^2$,
where $\eta = \sqrt{\langle{H}\rangle^2+\langle{H'}\rangle^2}$.
Those fields initially charged only under hypercharge now couple
in the same way as before to the $U(1)_Y$ multiplet, but they also
develop a coupling to the $U(1)_x$ multiplet of strength
\beq
-\tilde{g}_x\,s_{\epsilon}\lrf{g_Y}{\tilde{g_x}}Y_i \ .
\eeq
This shift induces a small amount of mass mixing among the vector and
gaugino fields.  For the vectors, the mixing is on the order of
$\epsilon^2m_Z^2/m_x^2$ while for the gauginos it is $\epsilon M_1/M_x$.
We neglect this effect since it is suppressed by both a small mass
ratio and the kinetic mixing parameter.

  Shifting the gauge multiplets also modifies the
hypercharge and $U(1)_x$ $D$ terms.  The $D$-term potential becomes
\bea
V_D &=& \frac{1}{2}c_{\epsilon}^2g_x^2\left(\sum_ix_i|H_i|^2\right)^2
+c_{\epsilon}^2(g_x^2\xi_x-\epsilon g_xg_Y\xi_Y)\sum_ix_i|H_i|^2
\label{dterm}
\\
&&+ \frac{1}{2}c_{\epsilon}^2g_Y^2\left(\sum_iY_i|\phi_i|^2\right)^2
+ c_{\epsilon}^2\left[g_Y^2\xi_Y
- \epsilon g_xg_Y(\sum_ix_i|H_i|^2+\xi_x)\right]\sum_iY_i|\phi_i|^2
\ .\nnmb
\eea
The terms quadratic in the scalar fields can be absorbed as shifts in
the soft squared masses.  From this, we see that kinetic mixing alters
the effective MSSM soft masses through both the $U(1)_x$ FI term
and the VEVs of $H$ and $H'$.  Using Eqs.~(\ref{firun},\ref{mtot}),
the explicit dependence on FI terms can be incorporated instead into
the running of the total scalar soft masses in both the visible and the $x$
sector.  This is the origin of the term proportional to
$S_x = x_H(m_H^2-m_{H'}^2)$ in Eq.~\eqref{misoft}~~\cite{Jack:2000nm}.
Incorporating the FI terms into the running of the $x$-sector
total soft masses, we find
\beq
(4\pi)^2\frac{dm_{H,H'}^2}{dt} =
-8 x_{H}^{2}g_{x}^{2}|M_{x}|^{2}
\pm 2 \tilde{g}_{x}^{2} x_{H}^2(m_{H}^{2}-m_{H'}^{2})
\mp 2s_{\epsilon} \sqrt{\frac{3}{5}}
x_{H} \tilde{g}_{x} g_{1} tr(Ym^{2}) \ .
\eeq
From Eq.~\eqref{dterm} we see as well that in integrating out the massive
$x$-sector at scale $m_{3/2}$, the effective MSSM
soft masses are shifted by
\beq
m_i^2 \to m_i^2 + \frac{s_{\epsilon}}{2}\,\frac{Y_i}{x_H}
\frac{g_Y}{\tilde{g}_x}\,m_x^2\cos2\alpha \ ,
\eeq
which coincides with the prescription of Eq.~\eqref{dshift}.


\section{Fine Tuning in the Scalar Potential}
\label{sec:ft}

  The $x$-sector scalar potential has a $D$-flat direction when the
two Higgs states have equal expectation values.  In this case, the potential
of Eq.~\eqref{eq:xpot} is purely quadratic and can not have a stable
non-zero minimum.
The requirement that the quadratic coefficient in the $D$-flat direction
is positive also disfavours a minimum with $\tan\alpha$ close to unity
when this parameter is determined from a uniform scan over the soft
supersymmetry breaking parameters in the $x$ sector;
it requires a fine tuning to make the combination
$m_H^2 + m_{H'}^2 + 2 \mu'{}^2 - 2 b$ much smaller than
the natural scale, $m_{3/2}^2$.
For this reason, in our parameter scans discussed in
Section~\ref{sec:viable} where we use a uniform prior on $\tan\alpha$,
we excise a small region around $\tan\alpha = 1$.

  To quantify this fine tuning, we perform a simple scan over
the Lagrangian parameters to answer two questions:
what is the implied distribution of $\tan\alpha$ for flat priors on
the $x$-sector supersymmetry breaking parameters;
and when do we get an ``unnatural'' cancellation between
the Lagrangian parameters, leading to very light states.
As the overall scale is unimportant, we treat the mass parameters
as dimensionless and use flat priors with the ranges given in
Table~\ref{tab:ftscan}.  Note that we may choose $b'$ positive without
loss of generality.  We randomly generate $10^6$ points,
rejecting those where the potential is not bounded from below
or the $U(1)_x$ gauge symmetry is not spontaneously broken.
For the survivors we calculate both $\tan\alpha$ and the mass
of the lightest bosonic state, the lighter real scalar.

\begin{table}
  \centering
  \begin{tabular}{|c|c|}
    \hline
    Parameter & Interval \\
    \hline
    $m_H^2$ & $[ -100, 100]$ \\
    $m_{H'}^2$ & $[ -100, 100]$ \\
    $b'$ & $[0,100]$ \\
    $\mu'$ & $[-10, 10]$ \\

    \hline
  \end{tabular}
  \caption{Scan range of Lagrangian parameters for exploration of fine tuning in the $x$ sector.  Note that the overall scale of these parameters is unimportant to this appendix.}\label{tab:ftscan}
\end{table}

\begin{figure}
  \includegraphics{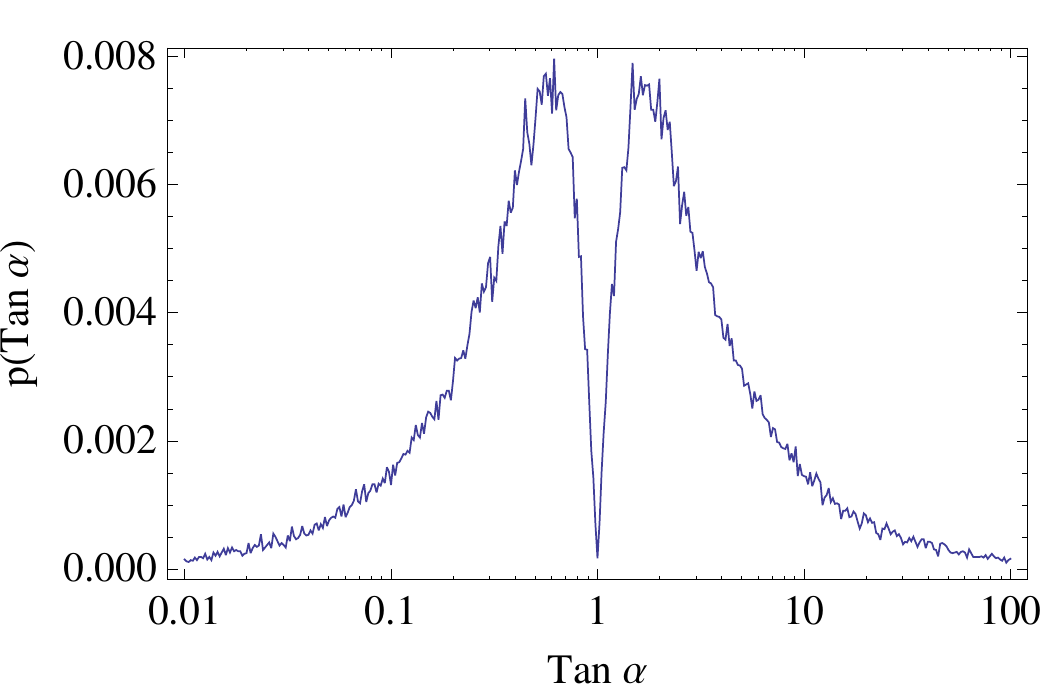}
  \caption{Distribution of $\tan\alpha$ implied by uniform priors on Lagrangian parameters, and imposing consistent spontaneous symmetry breaking.}\label{fig:tanadist}
\end{figure}

  We plot the (unnormalised) distribution of $\tan\alpha$
in Fig.~\ref{fig:tanadist}.  The general behaviour is easy
to understand.  Recall that $\alpha$ is given by
\begin{equation}
  \sin (2\alpha) = \frac{2b'}{m_H^2 + m_{H'}^2 + 2\mu'{}^2} .
\end{equation}
Uniform priors on the Lagrangian parameters would lead to an approximately
flat distribution in $\sin (2\alpha)$.  Demanding spontaneous symmetry
breaking as well imposes a lower limit on $b'$, and thus biases $\sin(2\alpha)$
-- and hence also $\tan\alpha$ -- towards unity.
The dip in Fig~\ref{fig:tanadist} \emph{at} $\tan\alpha = 1$ is due
to the Jacobian of the variable transformation:
\begin{equation}
  p (\tan\alpha) = \left\lvert \frac{d \sin(2\alpha)}{d \tan\alpha} \right\rvert \bar{p} (\sin(2\alpha) = \frac{\left\lvert 1 - \tan^2\alpha \right\rvert}{(1+\tan^2\alpha)^2} \; \bar{p} (\sin(2\alpha) .
\end{equation}
We see that $\tan\alpha$ very close to unity is thus theoretically disfavoured.

\begin{figure}
  \includegraphics{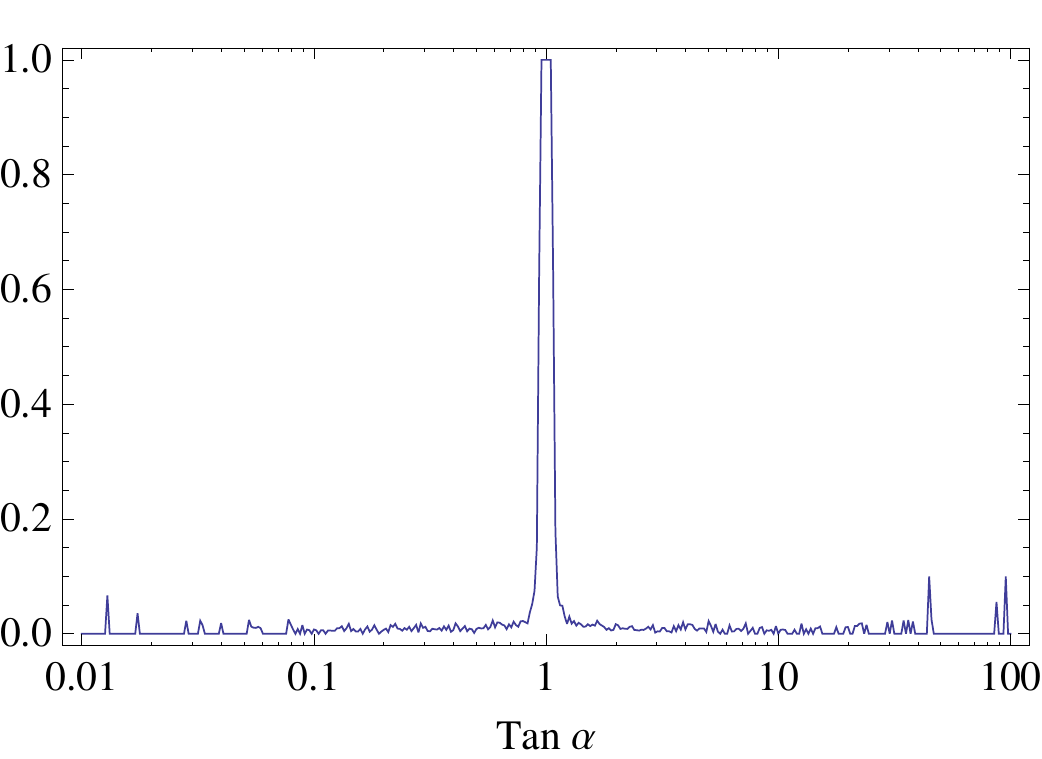}
  \caption{Fraction of points with scalars lighter than one-tenth
the mean Lagrangian parameters, as a function of $\tan\alpha$.}\label{fig:ltsca}
\end{figure}

Further, small values of $|\tan\alpha-1|$ are atypical in another sense.
In Fig.~\ref{fig:ltsca} we show the fraction of points for a given
$\tan\alpha$ for which the mass of the lightest scalar is one-tenth
or less of the average of the Lagrangian parameters:
\begin{equation}
  m_{h_{1x}}^2 \leq 10^{-2} \times \frac{1}{4} ( \abs{m_H^2} + \abs{m_{H'}^2} + \mu'{}^2 + b' ).
\end{equation}
This corresponds to a 1\%
cancellation that is unprotected by any symmetries.
It is clear that these points are almost exclusively associated with
$\tan\alpha$ close to 1.  They correspond to the near-flatness of the
potential for $|\tan\alpha - 1|\ll 1$, and they also coincide with
very large vector boson masses $m_x^2 \gg m_{3/2}^2$.
In our phenomenological analysis in Section~\ref{sec:viable} where
we scan over $\tan\alpha$ with a uniform prior, we exclude this unlikely
central peak by demanding $\left\lvert \tan\alpha - 1 \right\rvert > 0.1$.

\bibliographystyle{JHEP}
\bibliography{AMSBx}

\providecommand{\href}[2]{#2}\begingroup\raggedright\begin{thebibliography}{10}

\bibitem{Randall:1998uk}
L.~Randall and R.~Sundrum, {\it {Out of this world supersymmetry breaking}},
  {\em Nucl. Phys.} {\bf B557} (1999) 79--118,
  [\href{http://xxx.lanl.gov/abs/hep-th/9810155}{{\tt hep-th/9810155}}].

\bibitem{Giudice:1998xp}
G.~F. Giudice, M.~A. Luty, H.~Murayama, and R.~Rattazzi, {\it {Gaugino Mass
  without Singlets}},  {\em JHEP} {\bf 12} (1998) 027,
  [\href{http://xxx.lanl.gov/abs/hep-ph/9810442}{{\tt hep-ph/9810442}}].

\bibitem{Pomarol:1999ie}
A.~Pomarol and R.~Rattazzi, {\it {Sparticle masses from the superconformal
  anomaly}},  {\em JHEP} {\bf 05} (1999) 013,
  [\href{http://xxx.lanl.gov/abs/hep-ph/9903448}{{\tt hep-ph/9903448}}].

\bibitem{Katz:1999uw}
E.~Katz, Y.~Shadmi, and Y.~Shirman, {\it {Heavy thresholds, slepton masses and
  the mu term in anomaly mediated supersymmetry breaking}},  {\em JHEP} {\bf
  08} (1999) 015, [\href{http://xxx.lanl.gov/abs/hep-ph/9906296}{{\tt
  hep-ph/9906296}}].

\bibitem{Sundrum:2004un}
R.~Sundrum, {\it {'Gaugomaly' mediated SUSY breaking and conformal
  sequestering}},  {\em Phys. Rev.} {\bf D71} (2005) 085003,
  [\href{http://xxx.lanl.gov/abs/hep-th/0406012}{{\tt hep-th/0406012}}].

\bibitem{Cai:2010tj}
Y.~Cai and M.~A. Luty, {\it {Minimal Gaugomaly Mediation}},
  \href{http://xxx.lanl.gov/abs/1008.2024}{{\tt arXiv:1008.2024}}.

\bibitem{Kobayashi:2011bp}
T.~Kobayashi, Y.~Nakai, and M.~Sakai, {\it {(Extra)Ordinary Gauge/Anomaly
  Mediation}},  {\em JHEP} {\bf 1106} (2011) 039,
  [\href{http://xxx.lanl.gov/abs/1103.4912}{{\tt arXiv:1103.4912}}].

\bibitem{Okada:2002mv}
N.~Okada, {\it {Positively deflected anomaly mediation}},  {\em Phys.Rev.} {\bf
  D65} (2002) 115009, [\href{http://xxx.lanl.gov/abs/hep-ph/0202219}{{\tt
  hep-ph/0202219}}].

\bibitem{Jack:2000cd}
I.~Jack and D.~Jones, {\it {Fayet-Iliopoulos D terms and anomaly mediated
  supersymmetry breaking}},  {\em Phys.Lett.} {\bf B482} (2000) 167--173,
  [\href{http://xxx.lanl.gov/abs/hep-ph/0003081}{{\tt hep-ph/0003081}}].

\bibitem{ArkaniHamed:2000xj}
N.~Arkani-Hamed, D.~E. Kaplan, H.~Murayama, and Y.~Nomura, {\it {Viable
  ultraviolet-insensitive supersymmetry breaking}},  {\em JHEP} {\bf 02} (2001)
  041, [\href{http://xxx.lanl.gov/abs/hep-ph/0012103}{{\tt hep-ph/0012103}}].

\bibitem{Kaplan:2000jz}
D.~E. Kaplan and G.~D. Kribs, {\it {Gaugino-assisted anomaly mediation}},  {\em
  JHEP} {\bf 0009} (2000) 048,
  [\href{http://xxx.lanl.gov/abs/hep-ph/0009195}{{\tt hep-ph/0009195}}].

\bibitem{Chacko:2001jt}
Z.~Chacko and M.~A. Luty, {\it {Realistic anomaly mediation with bulk gauge
  fields}},  {\em JHEP} {\bf 0205} (2002) 047,
  [\href{http://xxx.lanl.gov/abs/hep-ph/0112172}{{\tt hep-ph/0112172}}].

\bibitem{Dermisek:2007qi}
R.~Dermisek, H.~Verlinde, and L.-T. Wang, {\it {Hypercharged Anomaly
  Mediation}},  {\em Phys.Rev.Lett.} {\bf 100} (2008) 131804,
  [\href{http://xxx.lanl.gov/abs/0711.3211}{{\tt arXiv:0711.3211}}].

\bibitem{deBlas:2009vx}
J.~de~Blas, P.~Langacker, G.~Paz, and L.-T. Wang, {\it {Combining Anomaly and
  Z-prime Mediation of Supersymmetry Breaking}},  {\em JHEP} {\bf 1001} (2010)
  037, [\href{http://xxx.lanl.gov/abs/0911.1996}{{\tt arXiv:0911.1996}}].

\bibitem{Chacko:1999am}
Z.~Chacko, M.~A. Luty, I.~Maksymyk, and E.~Ponton, {\it {Realistic anomaly
  mediated supersymmetry breaking}},  {\em JHEP} {\bf 0004} (2000) 001,
  [\href{http://xxx.lanl.gov/abs/hep-ph/9905390}{{\tt hep-ph/9905390}}].

\bibitem{Mohapatra:2007xq}
R.~N. Mohapatra, N.~Setzer, and S.~Spinner, {\it {Minimal Seesaw as an
  Ultraviolet Insensitive Cure for the Problems of Anomaly Mediation}},  {\em
  Phys. Rev.} {\bf D77} (2008) 053013,
  [\href{http://xxx.lanl.gov/abs/0707.0020}{{\tt arXiv:0707.0020}}].

\bibitem{Choi:2005ge}
K.~Choi, A.~Falkowski, H.~P. Nilles, and M.~Olechowski, {\it {Soft
  supersymmetry breaking in KKLT flux compactification}},  {\em Nucl.Phys.}
  {\bf B718} (2005) 113--133,
  [\href{http://xxx.lanl.gov/abs/hep-th/0503216}{{\tt hep-th/0503216}}].

\bibitem{Endo:2005uy}
M.~Endo, M.~Yamaguchi, and K.~Yoshioka, {\it {A Bottom-up approach to moduli
  dynamics in heavy gravitino scenario: Superpotential, soft terms and
  sparticle mass spectrum}},  {\em Phys.Rev.} {\bf D72} (2005) 015004,
  [\href{http://xxx.lanl.gov/abs/hep-ph/0504036}{{\tt hep-ph/0504036}}].

\bibitem{Acharya:2007rc}
B.~S. Acharya, K.~Bobkov, G.~L. Kane, P.~Kumar, and J.~Shao, {\it {Explaining
  the Electroweak Scale and Stabilizing Moduli in M Theory}},  {\em Phys.Rev.}
  {\bf D76} (2007) 126010, [\href{http://xxx.lanl.gov/abs/hep-th/0701034}{{\tt
  hep-th/0701034}}].

\bibitem{Luty:1999cz}
M.~A. Luty and R.~Sundrum, {\it {Radius stabilization and anomaly-mediated
  supersymmetry breaking}},  {\em Phys. Rev.} {\bf D62} (2000) 035008,
  [\href{http://xxx.lanl.gov/abs/hep-th/9910202}{{\tt hep-th/9910202}}].

\bibitem{Kachru:2007xp}
S.~Kachru, L.~McAllister, and R.~Sundrum, {\it {Sequestering in String
  Theory}},  {\em JHEP} {\bf 0710} (2007) 013,
  [\href{http://xxx.lanl.gov/abs/hep-th/0703105}{{\tt hep-th/0703105}}]. *
  Temporary entry *.

\bibitem{Luty:2001zv}
M.~Luty and R.~Sundrum, {\it {Anomaly mediated supersymmetry breaking in
  four-dimensions, naturally}},  {\em Phys.Rev.} {\bf D67} (2003) 045007,
  [\href{http://xxx.lanl.gov/abs/hep-th/0111231}{{\tt hep-th/0111231}}].

\bibitem{Schmaltz:2006qs}
M.~Schmaltz and R.~Sundrum, {\it {Conformal Sequestering Simplified}},  {\em
  JHEP} {\bf 0611} (2006) 011,
  [\href{http://xxx.lanl.gov/abs/hep-th/0608051}{{\tt hep-th/0608051}}].

\bibitem{Holdom:1985ag}
B.~Holdom, {\it {Two U(1)'s and Epsilon Charge Shifts}},  {\em Phys. Lett.}
  {\bf B166} (1986) 196.

\bibitem{Giudice:1988yz}
G.~F. Giudice and A.~Masiero, {\it {A Natural Solution to the mu Problem in
  Supergravity Theories}},  {\em Phys. Lett.} {\bf B206} (1988) 480--484.

\bibitem{Gherghetta:1999sw}
T.~Gherghetta, G.~F. Giudice, and J.~D. Wells, {\it {Phenomenological
  consequences of supersymmetry with anomaly induced masses}},  {\em
  Nucl.Phys.} {\bf B559} (1999) 27--47,
  [\href{http://xxx.lanl.gov/abs/hep-ph/9904378}{{\tt hep-ph/9904378}}].

\bibitem{Kaplan:1999ac}
D.~Kaplan, G.~D. Kribs, and M.~Schmaltz, {\it {Supersymmetry breaking through
  transparent extra dimensions}},  {\em Phys.Rev.} {\bf D62} (2000) 035010,
  [\href{http://xxx.lanl.gov/abs/hep-ph/9911293}{{\tt hep-ph/9911293}}].

\bibitem{Chacko:1999mi}
Z.~Chacko, M.~A. Luty, A.~E. Nelson, and E.~Ponton, {\it {Gaugino mediated
  supersymmetry breaking}},  {\em JHEP} {\bf 0001} (2000) 003,
  [\href{http://xxx.lanl.gov/abs/hep-ph/9911323}{{\tt hep-ph/9911323}}].

\bibitem{Fayet:1974jb}
P.~Fayet and J.~Iliopoulos, {\it {Spontaneously Broken Supergauge Symmetries
  and Goldstone Spinors}},  {\em Phys.Lett.} {\bf B51} (1974) 461--464.

\bibitem{Djouadi:2002ze}
A.~Djouadi, J.-L. Kneur, and G.~Moultaka, {\it {SuSpect: A Fortran code for the
  supersymmetric and Higgs particle spectrum in the MSSM}},  {\em
  Comput.Phys.Commun.} {\bf 176} (2007) 426--455,
  [\href{http://xxx.lanl.gov/abs/hep-ph/0211331}{{\tt hep-ph/0211331}}].

\bibitem{Heister:2002mn}
{\bf ALEPH} Collaboration, A.~Heister {\em et.~al.}, {\it {Search for charginos
  nearly mass degenerate with the lightest neutralino in e+ e- collisions at
  center-of-mass energies up to 209-GeV}},  {\em Phys.Lett.} {\bf B533} (2002)
  223--236, [\href{http://xxx.lanl.gov/abs/hep-ex/0203020}{{\tt
  hep-ex/0203020}}].

\bibitem{Abbiendi:2003sc}
{\bf OPAL} Collaboration, G.~Abbiendi {\em et.~al.}, {\it {Search for chargino
  and neutralino production at s**(1/2) = 192-GeV to 209 GeV at LEP}},  {\em
  Eur.Phys.J.} {\bf C35} (2004) 1--20,
  [\href{http://xxx.lanl.gov/abs/hep-ex/0401026}{{\tt hep-ex/0401026}}].

\bibitem{Batra:2003nj}
P.~Batra, A.~Delgado, D.~E. Kaplan, and T.~M. Tait, {\it {The Higgs mass bound
  in gauge extensions of the minimal supersymmetric standard model}},  {\em
  JHEP} {\bf 0402} (2004) 043,
  [\href{http://xxx.lanl.gov/abs/hep-ph/0309149}{{\tt hep-ph/0309149}}].

\bibitem{Barger:2006dh}
V.~Barger, P.~Langacker, H.-S. Lee, and G.~Shaughnessy, {\it {Higgs Sector in
  Extensions of the MSSM}},  {\em Phys.Rev.} {\bf D73} (2006) 115010,
  [\href{http://xxx.lanl.gov/abs/hep-ph/0603247}{{\tt hep-ph/0603247}}].

\bibitem{Martin:2009bg}
S.~P. Martin, {\it {Extra vector-like matter and the lightest Higgs scalar
  boson mass in low-energy supersymmetry}},  {\em Phys.Rev.} {\bf D81} (2010)
  035004, [\href{http://xxx.lanl.gov/abs/0910.2732}{{\tt arXiv:0910.2732}}].

\bibitem{Jack:2000nm}
I.~Jack, D.~Jones, and S.~Parsons, {\it {The Fayet-Iliopoulos D term and its
  renormalization in softly broken supersymmetric theories}},  {\em Phys.Rev.}
  {\bf D62} (2000) 125022, [\href{http://xxx.lanl.gov/abs/hep-ph/0007291}{{\tt
  hep-ph/0007291}}].

\bibitem{Rattazzi:1999qg}
R.~Rattazzi, A.~Strumia, and J.~D. Wells, {\it {Phenomenology of deflected
  anomaly mediation}},  {\em Nucl.Phys.} {\bf B576} (2000) 3--28,
  [\href{http://xxx.lanl.gov/abs/hep-ph/9912390}{{\tt hep-ph/9912390}}].

\bibitem{Baer:2000bs}
H.~Baer, J.~Mizukoshi, and X.~Tata, {\it {Reach of the CERN LHC for the minimal
  anomaly mediated SUSY breaking model}},  {\em Phys.Lett.} {\bf B488} (2000)
  367--372, [\href{http://xxx.lanl.gov/abs/hep-ph/0007073}{{\tt
  hep-ph/0007073}}].

\bibitem{Barr:2002ex}
A.~Barr, C.~Lester, M.~A. Parker, B.~Allanach, and P.~Richardson, {\it
  {Discovering anomaly mediated supersymmetry at the LHC}},  {\em JHEP} {\bf
  0303} (2003) 045, [\href{http://xxx.lanl.gov/abs/hep-ph/0208214}{{\tt
  hep-ph/0208214}}].

\bibitem{Baer:2010kd}
H.~Baer, R.~Dermisek, S.~Rajagopalan, and H.~Summy, {\it {Neutralino, axion and
  axino cold dark matter in minimal, hypercharged and gaugino AMSB}},  {\em
  JCAP} {\bf 1007} (2010) 014, [\href{http://xxx.lanl.gov/abs/1004.3297}{{\tt
  arXiv:1004.3297}}].

\bibitem{Gunion:2001fu}
J.~F. Gunion and S.~Mrenna, {\it {Probing models with near degeneracy of the
  chargino and LSP at a linear e+ e- collider}},  {\em Phys.Rev.} {\bf D64}
  (2001) 075002, [\href{http://xxx.lanl.gov/abs/hep-ph/0103167}{{\tt
  hep-ph/0103167}}].

\bibitem{Ibe:2006de}
M.~Ibe, T.~Moroi, and T.~Yanagida, {\it {Possible Signals of Wino LSP at the
  Large Hadron Collider}},  {\em Phys.Lett.} {\bf B644} (2007) 355--360,
  [\href{http://xxx.lanl.gov/abs/hep-ph/0610277}{{\tt hep-ph/0610277}}].

\bibitem{Thomas:1998wy}
S.~D. Thomas and J.~D. Wells, {\it {Phenomenology of Massive Vectorlike Doublet
  Leptons}},  {\em Phys.Rev.Lett.} {\bf 81} (1998) 34--37,
  [\href{http://xxx.lanl.gov/abs/hep-ph/9804359}{{\tt hep-ph/9804359}}].

\bibitem{Buckley:2009kv}
M.~R. Buckley, L.~Randall, and B.~Shuve, {\it {LHC Searches for Non-Chiral
  Weakly Charged Multiplets}},  \href{http://xxx.lanl.gov/abs/0909.4549}{{\tt
  arXiv:0909.4549}}.

\bibitem{Pierce:1996zz}
D.~M. Pierce, J.~A. Bagger, K.~T. Matchev, and R.-j. Zhang, {\it {Precision
  corrections in the minimal supersymmetric standard model}},  {\em Nucl.Phys.}
  {\bf B491} (1997) 3--67, [\href{http://xxx.lanl.gov/abs/hep-ph/9606211}{{\tt
  hep-ph/9606211}}].

\bibitem{Chen:1995yu}
C.~Chen, M.~Drees, and J.~Gunion, {\it {Searching for invisible and almost
  invisible particles at e+ e- colliders}},  {\em Phys.Rev.Lett.} {\bf 76}
  (1996) 2002--2005, [\href{http://xxx.lanl.gov/abs/hep-ph/9512230}{{\tt
  hep-ph/9512230}}]. Addendum/Erratum in hep-ph 9902309.

\bibitem{Djouadi:2006bz}
A.~Djouadi, M.~Muhlleitner, and M.~Spira, {\it {Decays of supersymmetric
  particles: The Program SUSY-HIT (SUspect-SdecaY-Hdecay-InTerface)}},  {\em
  Acta Phys.Polon.} {\bf B38} (2007) 635--644,
  [\href{http://xxx.lanl.gov/abs/hep-ph/0609292}{{\tt hep-ph/0609292}}].

\bibitem{Sjostrand:2006za}
T.~Sjostrand, S.~Mrenna, and P.~Z. Skands, {\it {PYTHIA 6.4 Physics and
  Manual}},  {\em JHEP} {\bf 0605} (2006) 026,
  [\href{http://xxx.lanl.gov/abs/hep-ph/0603175}{{\tt hep-ph/0603175}}].

\bibitem{Meade:2007js}
P.~Meade and M.~Reece, {\it {BRIDGE: Branching ratio inquiry / decay generated
  events}},  \href{http://xxx.lanl.gov/abs/hep-ph/0703031}{{\tt
  hep-ph/0703031}}.

\bibitem{Alwall:2007st}
J.~Alwall, P.~Demin, S.~de~Visscher, R.~Frederix, M.~Herquet, {\em et.~al.},
  {\it {MadGraph/MadEvent v4: The New Web Generation}},  {\em JHEP} {\bf 0709}
  (2007) 028, [\href{http://xxx.lanl.gov/abs/0706.2334}{{\tt
  arXiv:0706.2334}}].

\bibitem{atlas-conf-2011-086}
{\bf ATLAS} Collaboration, ATLAS-CONF-2011-086, ``{SUSY Search with jets and
  Etmiss}.''.

\bibitem{cms-pas-sus-11-003}
{\bf CMS} Collaboration, CMS-PAS-SUS-11-003, ``{Search for supersymmetry in
  all-hadronic events with alphaT }.''.

\bibitem{cms-pas-sus-11-004}
{\bf CMS} Collaboration, CMS-PAS-SUS-11-004, ``{Search for supersymmetry in
  all-hadronic events with missing energy }.''.

\bibitem{cms-pas-sus-11-005}
{\bf CMS} Collaboration, CMS-PAS-SUS-11-005, ``{Search for supersymmetry in
  all-hadronic events with MT2 }.''.

\bibitem{atlas-conf-2011-098}
{\bf ATLAS} Collaboration, ATLAS-CONF-2011-098, ``{SUSY Search with bjets and
  Etmiss}.''.

\bibitem{Abazov:2010wq}
{\bf D0} Collaboration, V.~M. Abazov {\em et.~al.}, {\it {Search for scalar
  bottom quarks and third-generation leptoquarks in p p-bar collisions at
  sqrt(s) = 1.96 TeV}},  {\em Phys.Lett.} {\bf B693} (2010) 95--101,
  [\href{http://xxx.lanl.gov/abs/1005.2222}{{\tt arXiv:1005.2222}}].

\bibitem{Aaltonen:2010dy}
{\bf CDF} Collaboration, T.~Aaltonen {\em et.~al.}, {\it {Search for the
  Production of Scalar Bottom Quarks in $p \bar {p} $ collisions at $\sqrt{s} $
  = 1.96 TeV}},  {\em Phys.Rev.Lett.} {\bf 105} (2010) 081802,
  [\href{http://xxx.lanl.gov/abs/1005.3600}{{\tt arXiv:1005.3600}}].

\bibitem{ArkaniHamed:2006mb}
N.~Arkani-Hamed, A.~Delgado, and G.~Giudice, {\it {The Well-tempered
  neutralino}},  {\em Nucl.Phys.} {\bf B741} (2006) 108--130,
  [\href{http://xxx.lanl.gov/abs/hep-ph/0601041}{{\tt hep-ph/0601041}}].

\bibitem{Moroi:1999zb}
T.~Moroi and L.~Randall, {\it {Wino cold dark matter from anomaly mediated SUSY
  breaking}},  {\em Nucl.Phys.} {\bf B570} (2000) 455--472,
  [\href{http://xxx.lanl.gov/abs/hep-ph/9906527}{{\tt hep-ph/9906527}}].

\bibitem{Acharya:2009zt}
B.~S. Acharya, G.~Kane, S.~Watson, and P.~Kumar, {\it {A Non-thermal WIMP
  Miracle}},  {\em Phys.Rev.} {\bf D80} (2009) 083529,
  [\href{http://xxx.lanl.gov/abs/0908.2430}{{\tt arXiv:0908.2430}}].

\bibitem{Grajek:2008pg}
P.~Grajek, G.~Kane, D.~Phalen, A.~Pierce, and S.~Watson, {\it {Is the PAMELA
  Positron Excess Winos?}},  {\em Phys.Rev.} {\bf D79} (2009) 043506,
  [\href{http://xxx.lanl.gov/abs/0812.4555}{{\tt arXiv:0812.4555}}].

\bibitem{Kane:2009if}
G.~Kane, R.~Lu, and S.~Watson, {\it {PAMELA Satellite Data as a Signal of
  Non-Thermal Wino LSP Dark Matter}},  {\em Phys.Lett.} {\bf B681} (2009)
  151--160, [\href{http://xxx.lanl.gov/abs/0906.4765}{{\tt arXiv:0906.4765}}].
  * Brief entry *.

\bibitem{Adriani:2008zr}
{\bf PAMELA} Collaboration, O.~Adriani {\em et.~al.}, {\it {An anomalous
  positron abundance in cosmic rays with energies 1.5-100 GeV}},  {\em Nature}
  {\bf 458} (2009) 607--609, [\href{http://xxx.lanl.gov/abs/0810.4995}{{\tt
  arXiv:0810.4995}}].

\bibitem{Adriani:2010rc}
{\bf PAMELA} Collaboration, O.~Adriani {\em et.~al.}, {\it {PAMELA results on
  the cosmic-ray antiproton flux from 60 MeV to 180 GeV in kinetic energy}},
  {\em Phys.Rev.Lett.} {\bf 105} (2010) 121101,
  [\href{http://xxx.lanl.gov/abs/1007.0821}{{\tt arXiv:1007.0821}}].

\bibitem{Abdo:2010nc}
A.~Abdo, M.~Ackermann, M.~Ajello, W.~Atwood, L.~Baldini, {\em et.~al.}, {\it
  {Fermi LAT Search for Photon Lines from 30 to 200 GeV and Dark Matter
  Implications}},  {\em Phys.Rev.Lett.} {\bf 104} (2010) 091302,
  [\href{http://xxx.lanl.gov/abs/1001.4836}{{\tt arXiv:1001.4836}}].

\bibitem{Abdo:2010ex}
A.~Abdo, M.~Ackermann, M.~Ajello, W.~Atwood, L.~Baldini, {\em et.~al.}, {\it
  {Observations of Milky Way Dwarf Spheroidal galaxies with the Fermi-LAT
  detector and constraints on Dark Matter models}},  {\em Astrophys.J.} {\bf
  712} (2010) 147--158, [\href{http://xxx.lanl.gov/abs/1001.4531}{{\tt
  arXiv:1001.4531}}]. * Temporary entry *.

\bibitem{Slatyer:2009yq}
T.~R. Slatyer, N.~Padmanabhan, and D.~P. Finkbeiner, {\it {CMB Constraints on
  WIMP Annihilation: Energy Absorption During the Recombination Epoch}},  {\em
  Phys.Rev.} {\bf D80} (2009) 043526,
  [\href{http://xxx.lanl.gov/abs/0906.1197}{{\tt arXiv:0906.1197}}].

\bibitem{Aprile:2011hi}
{\bf XENON100} Collaboration, E.~Aprile {\em et.~al.}, {\it {Dark Matter
  Results from 100 Live Days of XENON100 Data}},  {\em Phys.Rev.Lett.} (2011)
  [\href{http://xxx.lanl.gov/abs/1104.2549}{{\tt arXiv:1104.2549}}].

\bibitem{Giudice:1997ni}
G.~F. Giudice and R.~Rattazzi, {\it {Extracting Supersymmetry-Breaking Effects
  from Wave- Function Renormalization}},  {\em Nucl. Phys.} {\bf B511} (1998)
  25--44, [\href{http://xxx.lanl.gov/abs/hep-ph/9706540}{{\tt
  hep-ph/9706540}}].

\bibitem{ArkaniHamed:1998kj}
N.~Arkani-Hamed, G.~F. Giudice, M.~A. Luty, and R.~Rattazzi, {\it
  {Supersymmetry-breaking loops from analytic continuation into superspace}},
  {\em Phys. Rev.} {\bf D58} (1998) 115005,
  [\href{http://xxx.lanl.gov/abs/hep-ph/9803290}{{\tt hep-ph/9803290}}].

\bibitem{Morrissey:2009ur}
D.~E. Morrissey, D.~Poland, and K.~M. Zurek, {\it {Abelian Hidden Sectors at a
  GeV}},  {\em JHEP} {\bf 07} (2009) 050,
  [\href{http://xxx.lanl.gov/abs/0904.2567}{{\tt arXiv:0904.2567}}].

\bibitem{Martin:1997ns}
S.~P. Martin, {\it {A Supersymmetry primer}},
  \href{http://xxx.lanl.gov/abs/hep-ph/9709356}{{\tt hep-ph/9709356}}.

\end{thebibliography}\endgroup

\end{document}